\documentclass[showpacs,prd,twocolumn,aps,nofootinbib,superscriptaddress]{revtex4-1}

\usepackage{amsmath}
\usepackage{amssymb}
\usepackage{latexsym}
\usepackage{amsfonts}
\usepackage{epsfig}
\usepackage{psfrag}
\usepackage{graphicx}
\usepackage{wasysym}
\usepackage{ulem}
\usepackage{color}
\usepackage{slashed}
\usepackage{setspace}

\newcommand{\Eqref}[1]{Eq.~\eqref{#1}}
\renewcommand{\vec}[1]{\mathbf{#1}}

\begin{document}

\title{Quark-antiquark static energy from a restricted Fourier transform}

\author{Felix Karbstein}
\affiliation{Theoretisch-Physikalisches Institut,
Friedrich-Schiller-Universit\"at Jena, Max-Wien-Platz 1, D-07743 Jena, Germany}
\affiliation{Helmholtz-Institut Jena, Fr\"obelstieg 3, D-07743 Jena, Germany}

\begin{abstract}
 We provide a fully analytical determination of the perturbative quark-antiquark static energy in position space as defined by a restricted Fourier transformation from momentum to position space.
 Such a determination is complicated by the fact that the static energy genuinely decomposes into a strictly perturbative part (made up of contributions $\sim\alpha_s^n$, with $n\in\mathbb{N}$) which is conventionally evaluated in momentum space,
 and a so-called ultrasoft part (including terms $\sim\alpha_s^{n+m}\ln^m\alpha_s$, with  $n\geq3$ and $m\in\mathbb{N}$) which, conversely, is naturally evaluated in position space.
 Our approach facilitates the explicit determination of the static energy in position space at the accuracy with which the perturbative potential in momentum space is known, i.e., presently up to order $\alpha_s^4$. 
\end{abstract}

\date{\today}


\maketitle

\section{Introduction}

The interaction energy $E(r)$ of the color-singlet state made up of a static quark $Q$ and a static antiquark $\bar Q$ separated by a distance $r=|\vec{r}|$, the so-called $Q\bar Q$ static energy, constitutes one of the most basic quantities of quantum chromodynamics (QCD) and has been studied from its early days. It is also relevant for various phenomenological applications, like, e.g., quarkonium spectroscopy.
It is defined by \cite{Brown:1979ya}
\begin{equation}
 E(r)=\lim_{T\to\infty}\frac{i}{T}\ln\langle W_\square(r,T)\rangle\,,
\end{equation}
where $\langle W_\square(r,T)\rangle$ denotes the rectangular static Wilson loop \cite{Wilson:1974sk}.
As $\langle W_\square(r,T)\rangle$ can straightforwardly be evaluated on the lattice, in the parameter regime accessible by lattice simulations $E(r)$ can directly be determined in position space. Conversely, in the manifestly perturbative regime, the situation is completely different: Higher-order calculations in perturbative quantum field theory are most conveniently performed in momentum space. In QCD perturbative calculations are viable at large momentum transfers due to the fact that QCD is asymptotically free. Correspondingly, the strictly perturbative contributions to the $Q\bar Q$ static energy, constituting the {\it perturbative potential} have been derived in momentum space and for large momentum transfers, $p=|\vec{p}|>\Lambda_{\rm QCD}$.

We denote this strictly perturbative potential in momentum space by $\tilde V(p)$.\footnote{In our notations quantities in position space are labeled by ordinary characters, e.g., $E$, while their momentum-space analogues are labeled by the same character denoted with a tilde, e.g., $\tilde E$.
In addition, we use calligraphic letters, e.g., $\cal E$, in statements that hold both in position and momentum space.} 
In standard perturbation theory loop diagrams come along with integrations $\int\frac{{\rm d}^4q}{(2\pi)^4}$ of the loop four-momentum $q$ over the full momentum regime.
Hence, such loops naturally also receive contributions from momenta $\lesssim\Lambda_{\rm QCD}$ for which perturbation theory is no longer trustworthy.
The leading uncontrolled contribution contained in $\tilde V(p)$ arising from this kind of diagrams is quadratic in $\Lambda_{\rm QCD}$ and $\sim-\frac{4\pi}{p^2}\alpha_s(\frac{\Lambda_{\rm QCD}}{p})^2$ \cite{Beneke:1998rk}.

$\tilde V(p)$ is presently known up to ${\cal O}(\alpha_s^4)$ accuracy \cite{Anzai:2009tm,Smirnov:2009fh}.
Noteworthily at ${\cal O}(\alpha_s^4)$ it features an explicit dependence on an {\it ultrasoft} (US) momentum scale $\mu_{\rm us}$ \cite{Appelquist:1977es,Brambilla:1999qa}.

However, as has been realized already long ago \cite{Appelquist:1977es}, the $Q\bar Q$ static energy does not have a strict power-series expansion in $\alpha_s$. Beyond ${\cal O}(\alpha_s^3)$, also logarithmic contributions in $\alpha_s$ are induced.
The difference between the static energy ${\cal E}$ and the static potential ${\cal V}$ are encoded in {\it ultrasoft corrections} ${\cal V}^{\rm US}$, which include terms $\sim\alpha_s^{n+m}\ln^m\alpha_s$, with  $n\geq3$ and $m\in\mathbb{N}$ \cite{Brambilla:2006wp,Brambilla:2009bi}.
These US corrections genuinely also depend explicitly on $\mu_{\rm us}$. More specifically, the dependence is such that the individual $\mu_{\rm us}$ dependences of $\cal V$ and ${\cal V}^{\rm US}$ exactly cancel, when adding them to form the $Q\bar Q$ singlet static energy in the perturbative regime.
Correspondingly, the static energy at ${\cal O}(\alpha_s^4)$ is rendered manifestly independent of any auxiliary scale (cf. also Secs.~\ref{seq:basics} and \ref{seq:statEresFT} below).
Schematically, it is thus given by
\begin{equation}
 {\cal E}={\cal V}(\mu_{\rm us})+{\cal V}^{\rm US}(\mu_{\rm us})\,. \label{eq:calE}
\end{equation}
To keep notations compact, we will subsequently often omit the explicit reference to the scale $\mu_{\rm us}$ in the argument of $\cal V$ and ${\cal V}^{\rm US}$.

The situation is further complicated by the fact that while $\cal V$ is directly accessible in momentum space, the US corrections ${\cal V}^{\rm US}$ in turn are most conveniently evaluated in position space.

To allow for insights into the perturbative energy in position space, the momentum space potential $\tilde V(p)$ has to be transformed into position space.
Even though it seems to be straightforwardly achievable by means of a standard Fourier transform on first sight, this step turns out to be highly nontrivial. It gives rise to a dilemma: A Fourier transform naturally requires information about the function to be transformed in the full momentum regime.
The perturbative expressions, however, are manifestly limited to a certain momentum regime; in QCD to $p\gg\Lambda_{\rm QCD}$, such that $\alpha_s(p)\ll1$.

As $\tilde V(p)$ is not at all trustworthy at low momenta, the low momentum part of the Fourier integral,
\begin{equation}
 V(r) = \int\frac{{\rm d}^3p}{(2\pi)^3}\,{\rm e}^{i\vec{p}\cdot\vec{r}}\,\tilde V(p)\,, \label{eq:V(r)}
\end{equation}
generically induces uncontrolled contributions, the leading one being linear in $\Lambda_{\rm QCD}$ and $\sim\frac{1}{r}\alpha_s(r\Lambda_{\rm QCD})$ \cite{Beneke:1998rk} (cf. also below).
In turn, $V(r)$ as defined by \Eqref{eq:V(r)} is more sensitive to long distances than $\tilde V(p)$.

This suggests to introduce a momentum cutoff $\mu_f>\Lambda_{\rm QCD}$ and to define the static potential in position space by means of a restricted Fourier transform as \cite{Beneke:1998rk}
\begin{align}
 V(r,\mu_f) &= \int_{|\vec{p}|>\mu_f}\frac{{\rm d}^3p}{(2\pi)^3}\,{\rm e}^{i\vec{p}\cdot\vec{r}}\,\tilde V(p) \nonumber\\
            &= V(r)-\delta V(r,\mu_f)\,, \label{eq:resFT}
\end{align}
with low momentum part
\begin{equation}
 \delta V(r,\mu_f) = \int_{|\vec{p}|<\mu_f}\frac{{\rm d}^3p}{(2\pi)^3}\,{\rm e}^{i\vec{p}\cdot\vec{r}}\,\tilde V(p)\,. \label{eq:deltaV}
\end{equation}
Obviously $\delta V$ vanishes for $\mu_f=0$, such that $V(r,\mu_f=0)=V(r)$ reproduces the result of an ordinary, unrestricted Fourier transform.

Beneke \cite{Beneke:1998rk} employed the decomposition~\eqref{eq:resFT} to propose the potential-subtracted (PS) scheme on its basis:
Aiming at the subtraction of the leading uncontrolled contribution of order $\Lambda_{\rm QCD}$ in $V(r)$, he makes use of $\delta V(r,\mu_f)=\delta V(\mu_f)+\mu_f{\cal O}(r^2\mu_f^2)$, where $\delta V(\mu_f)\equiv\delta V(r=0,\mu_f)\sim\mu_f$, and completely neglects the corrections of relative order $r^2\mu_f^2$ in \Eqref{eq:deltaV} to define the subtraction term in the PS scheme, which thus becomes independent of $r$ and reads $\delta V(\mu_f) = \int_{|\vec{p}|<\mu_f}\!\!\frac{{\rm d}^3p}{(2\pi)^3}\,\tilde V(p)$. Correspondingly, the subtraction term $\delta V(\mu_f)$ naturally has an expansion in $\alpha_s$, and is explicitly know with the accuracy of $\tilde V(p)$.
Due to the fact that no contributions of this order are present already in $\tilde V(p)$, in this way the contribution $\sim\Lambda_{\rm QCD}$ can be subtracted completely.
It can even be shown that this contribution exactly cancels against an analogous contribution in the pole mass \cite{Beneke:1998rk}.
Beyond linear order in $\Lambda_{\rm QCD}$ the situation becomes more intricate:
Not only the Fourier transform induces uncontrolled contributions, but -- as mentioned before -- there are also ones already present in $\tilde V(p)$ as determined in standard perturbation theory.

However, note that at the time when Beneke wrote his paper \cite{Beneke:1998rk}, the static potential was only known up to ${\cal O}(\alpha_s^3)$. Thus, nontrivial questions like the consistent adaption of the PS scheme to US corrections, ensuring the overall $\mu_{\rm us}$ independence of the static energy in the PS scheme were not addressed. As the approach pursued in our paper can be reduced to the PS schema by taking a certain limit (cf. below),
our paper effectively also tackles such questions. In this sense, as a by-product our paper provides the full, consistent expression of the $Q\bar Q$ static energy in the PS scheme at the presently best achievable accuracy, i.e., at ${\cal O}(\alpha_s^4)$.

Somewhat differently, but based on \cite{Beneke:1998rk}, Laschka et al. \cite{Laschka:2011zr} recently also employed \Eqref{eq:resFT} to define the perturbative potential in position space such that it does not suffer from the uncontrolled contribution $\sim\Lambda_{\rm QCD}$.
In contrast to \cite{Beneke:1998rk} they directly evaluate \Eqref{eq:resFT} numerically, using the full four-loop running of $\alpha_s(p)$ according to the renormalization group equation for $p>\mu_f$.

Other approaches to tackle these problems utilize a Borel transform of $V(r)$: The ambiguity contained in $V(r)$ is then attributed to the {\it renormalon poles} in the complex Borel plane \cite{Aglietti:1995tg}.
In particular, the ambiguity of order $\Lambda_{\rm QCD}$ follows from the pole closest to the origin \cite{Hoang:1998nz,Beneke:1998ui}. This observation also forms the basis of the renormalon subtracted (RS) scheme \cite{Pineda:2001zq,Pineda:2002se}

In this paper we provide an additional perspective.
Sticking to the $\overline{\rm MS}$-scheme we analytically evaluate \Eqref{eq:deltaV} for $p\geq\mu_f\gg\Lambda_{\rm QCD}$, without resorting to further approximations. The latter inequality ensures $\mu_f$ to be a perturbative momentum scale.
In close analogy to $\delta V(\mu_f)$ of \cite{Beneke:1998rk}, our result for $\delta V(r,\mu_f)$ has an expansion in $\alpha_s$, and is known explicitly at the accuracy of $\tilde V(p)$.
We in particular manage to analytically take into account the full $r$ dependence of \Eqref{eq:deltaV} at a given order of the expansion in $\alpha_s$.
Since $\delta V(r,\mu_f)=\delta V(\mu_f)+\mu_f{\cal O}(r^2\mu_f^2)$, with $\delta V(\mu_f)\sim\mu_f$ accounting for the entire contribution linear in $\Lambda_{\rm QCD}$,
the subtraction term in the {\it restricted Fourier transform scheme} advocated here differs from that in the PS scheme \cite{Beneke:1998rk} by terms of order $\mu_f{\cal O}(r^2\mu_f^2)\sim\Lambda_{\rm QCD}{\cal O}(r^2\Lambda_{\rm QCD}^2)$.
Hence, with respect to their order in $\Lambda_{\rm QCD}$ these terms are not more important than the uncontrolled ones genuinely contained in $\tilde V(p)$ (cf. the discussion before).

What motivates us to also subtract the higher order terms in \Eqref{eq:deltaV} is the observation that the Fourier transform~\eqref{eq:V(r)} does not only result in an uncontrolled contribution linear in $\Lambda_{\rm QCD}$, but also those $\sim\Lambda_{\rm QCD}{\cal O}(r^2\Lambda_{\rm QCD}^2)$ 
restrict the radius of convergence of an expansion of $V(r)$ in powers of $\alpha_s$:
As we will demonstrate explicitly up to ${\cal O}(\alpha_s^4)$ below, the coefficients of $\alpha_s^{1+k}$ with $k\in\mathbb{N}$ increase with $k$.
Even though this increase is more pronounced for the contribution $\sim\Lambda_{\rm QCD}$, it is also clearly visible for the contributions $\sim\Lambda_{\rm QCD}(r^2\Lambda_{\rm QCD}^2)$.
Closely examining the structure of the subtraction term $\delta V(r,\mu_f)$, we provide indications that an analogous increase is also to be expected for higher orders.
Correspondingly, by subtracting all these contributions in $V(r)$ by means of the restricted Fourier transform~\eqref{eq:restrictedFT} we may hope to further improve the convergence properties of the perturbative potential in position space.
Moreover, in this way we can at least assure that the transition from momentum to position space does not induce any new uncontrolled contributions not already present in $\tilde V(p)$.
In fact, given $\tilde V(p)$ at a certain accuracy in $\alpha_s$, we can quantify the contribution $\sim\mu_f(r^2\mu_f^2)^l$ to $V(r,\mu_f)$ for any given $l\in\mathbb{N}_0$, and evaluate the {\it infinite sum contribution} $\delta V(r,\mu_f)\sim\mu_f\sum_{l=0}^\infty(r^2\mu_f^2)^l$ originating in the Fourier transform over momenta where perturbation theory is no longer trustworthy.

Moreover, in a second step we manage to consistently adopt the restricted Fourier transform scheme to the ultrasoft corrections.
These considerations are also of relevance for the consistent adaption of generic {\it renormalon subtraction schemes} to the ultrasoft corrections.
As will be explained in detail in the subsequent sections -- even though they are directly evaluated in position space -- also the ultrasoft corrections receive an explicit $\mu_f$ dependence, such that the corresponding static energy in position space is given by
\begin{equation}
 E(r,\mu_f)=V(r,\mu_f)+V^{\rm US}(r,\mu_f)\,. \label{eq:E_rFT}
\end{equation}
Equation~\eqref{eq:E_rFT} generalizes the result obtained by a standard unrestricted Fourier transform,
\begin{equation}
 E(r)=V(r)+V^{\rm US}(r)\,, \label{eq:E_sFT}
\end{equation}
which amounts to the $\mu_f=0$ limit of \Eqref{eq:E_rFT}, i.e., $V^{\rm US}(r,\mu_f=0)=V^{\rm US}(r)$ and $E(r,\mu_f=0)=E(r)$.

Eventually also the full $Q\bar Q$ singlet static energy in position space can be decomposed as [cf. \Eqref{eq:resFT}]
\begin{equation}
 E(r,\mu_f)=E(r)-\delta E(r,\mu_f)\,,\label{eq:resFT-E}
\end{equation}
where $\delta E(r,\mu_f)$ encodes the differences between the $Q\bar Q$ singlet static energy as defined by a standard unrestricted Fourier transform and in the restricted Fourier transform scheme.
Most notably, this facilitates a complete, fully analytical determination of $E(r,\mu_f)$ at the accuracy with which $E(r)$ is known.

In order to further motivate the need of a subtraction scheme as provided by the restricted Fourier transform -- and to keep the paper self-contained --,
we show an exemplary plot of $E(r)$ for different accuracies in the perturbative expansion in Fig.~\ref{fig:E0_1234}:
\begin{figure}[h]
\center
\includegraphics[width=0.5\textwidth]{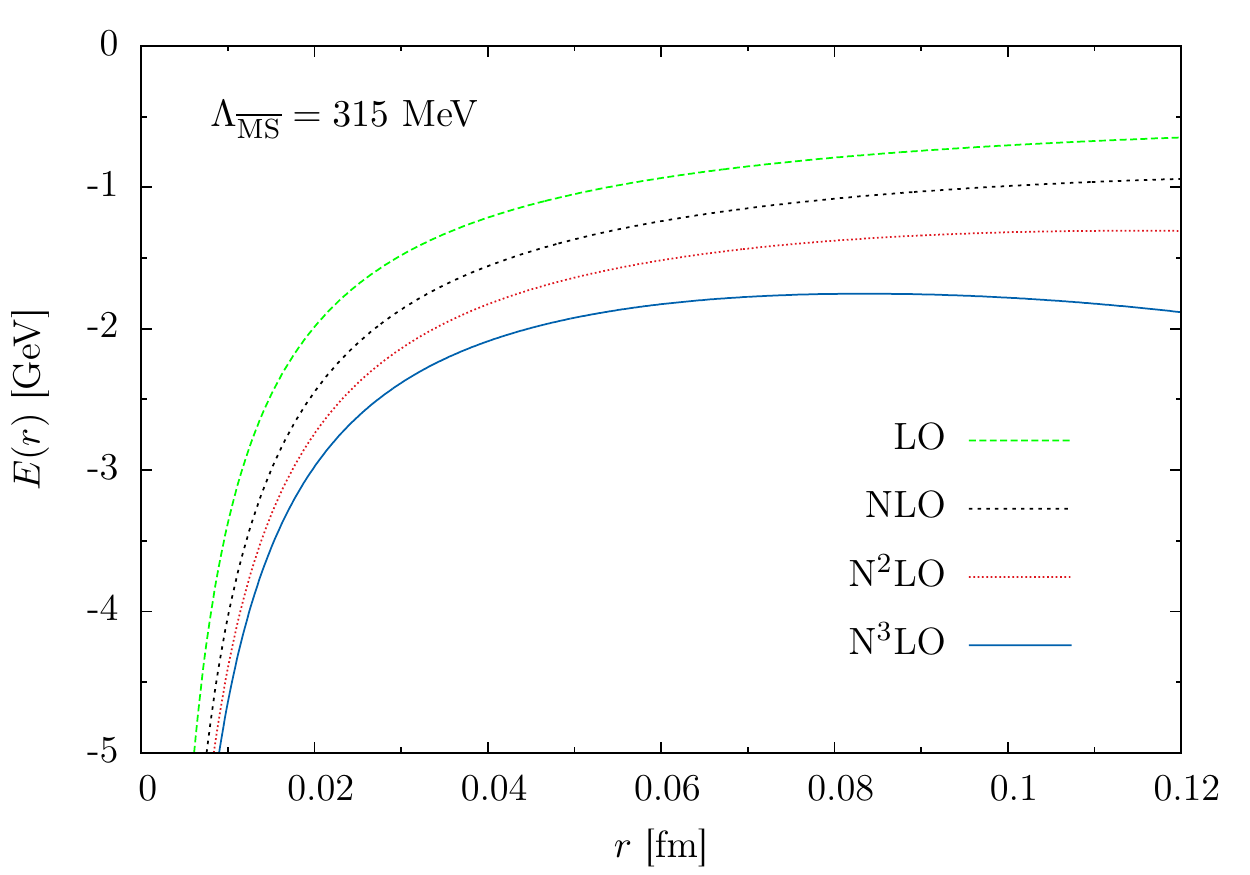}
\caption{The singlet static energy $E(r)$ in position space (for $n_f=2$) as obtained by a standard Fourier transform from momentum space.
To arrive at this plot we identify the renormalization scale $\mu$ with the inverse of the distance $r$ between the static quark and antiquark, $\mu=1/r$.
The accuracy of the perturbative static energy is gradually increased from leading order (LO) up to (next-to)$^3$LO ($\rm N^3LO$) accuracy, while
the 4-loop running of the coupling $\alpha_s$ is used throughout all expressions.
Over the depicted $r$ range, the coupling increases monotonically from $\alpha_s(1/r)|_{r=0}=0$ to $\alpha_s(1/r)|_{r=0.12{\rm fm}}\approx0.295$.
Given that $\alpha_s$ remains so small all over the $r$ range depicted here, the static energy behaves rather pathologically:
At least for $r\gtrsim0.06{\rm fm}$, the considered orders do not seem to converge in the sense that the corrections due to higher order contributions become increasingly less important. Conversely, the $\rm N^3LO$ result even bents back and decreases towards larger values of $r$. 
}
\label{fig:E0_1234}
\end{figure}
While we have not yet discussed any details necessary to arrive at this plot -- and correspondingly the reader is not expected to fully understand the specifications as provided in the caption of Fig.~\ref{fig:E0_1234} at this instance --, it should be clearly visible that the static energy~\eqref{eq:E_sFT} as defined by a standard Fourier transform behaves rather pathologically \cite{Billoire:1979ih}. At least for $r\gtrsim0.06{\rm fm}$, the considered orders do not seem to converge in the sense that the corrections due to higher order contributions become increasingly less important. The claim is that this pathological behavior primarily originates in the uncontrolled contributions at low momenta, to be cut out explicitly when resorting to a restricted Fourier transform. 
We will return to this plot in the {\it results and discussion} section (Sec.~\ref{seq:Ex+Res}) of this paper and confront it with the corresponding plot for \Eqref{eq:E_sFT}.

Our paper is organized as follows: After reviewing the present knowledge of the $Q\bar Q$ singlet static energy in Sec.~\ref{seq:basics}, in a way that it can straightforwardly be generalized to a restricted Fourier transform scheme also, we explicitly work out the various contributions to the perturbative $Q\bar Q$ singlet static energy as defined by a restricted Fourier transform from momentum to position space in Sec.~\ref{seq:statEresFT}.
In Sec.~\ref{seq:Ex+Res} we discuss some exemplary results, specializing to $n_f=2$ ($\Lambda_{\overline{\rm MS}}=315{\rm MeV}$ \cite{Jansen:2011vv}) and $\mu_f\in[3\ldots7]\Lambda_{\overline{\rm MS}}$. Finally, we end with conclusions in Sec.~\ref{seq:Concl}.

\section{Basic facts about the static energy in the perturbative regime} \label{seq:basics}

The running of the coupling $\alpha_s(\mu)$ as a function of the renormalization scale $\mu$ is one of the most essential and important ingredients for perturbative calculations.
It is governed by the QCD $\beta$-function, defined as
\begin{equation}
 \beta[\alpha_s(\mu)]\equiv\frac{\mu}{\alpha_s(\mu)}\frac{d}{d \mu}\alpha_s(\mu)\,, \label{eq:beta}
\end{equation}
which has the following series expansion in powers of $\alpha_s(\mu)$,
\begin{equation}
 \beta[\alpha_s(\mu)]=-\frac{\alpha_s(\mu)}{2\pi}\sum_{n=0}^\infty\left(\frac{\alpha_s(\mu)}{4\pi}\right)^n\beta_n\,. \label{eq:betaseries}
\end{equation}
The expansion coefficients $\beta_n$ are known up to $n=3$, i.e., to 4-loop order.
While $\beta_0$ and $\beta_1$ are independent of the renormalization scheme,
$\beta_3$ and $\beta_4$ are scheme-dependent. They have been
determined for arbitrary compact semi-simple Lie groups in the $\overline{\rm
MS}$-scheme \cite{vanRitbergen:1997va}. 
For SU(3) they read
\begin{align}
 \beta_0 &= 11-\frac{2}{3}n_f\,, \quad \beta_1= 102-\frac{38}{3}n_f\,, \nonumber\\
 \beta_2 &= \frac{2857}{2}-\frac{5033}{18}n_f+\frac{325}{54}n_f^2\,, \nonumber\\
 \beta_3 &= \frac{149753}{6}+3564\zeta(3)-\left(\frac{1078361}{162}+\frac{6508}{27}\zeta(3)\right)n_f \nonumber\\ 
         &\quad+\left(\frac{50065}{162}+\frac{6472}{81}\zeta(3)\right)n_f^2 +\frac{1093}{729} n_f^3\,,
\end{align}
with $n_f$ denoting the number of massless dynamical quark flavors.

As $\cal E$ is a physical observable, it should of course be independent of the explicit value of the renormalization scale $\mu$
and form a renormalization group (RG) invariant, i.e., fulfill
\begin{equation}
 \mu\frac{d}{d\mu}{\cal E}=0\,. \label{eq:RG_V}
\end{equation}
Equation~\eqref{eq:RG_V} is also expected to hold order by order in a strictly perturbative expansion in powers of $\alpha_s$, in the sense that given the perturbative potential $\cal V$ up to ${\cal O}(\alpha_s^{\bar k})$, we have
\begin{equation}
 \mu\frac{d}{d\mu}{\cal V}={\cal O}(\alpha_s^{\bar k+1})\,. \label{eq:RG_Vpert}
\end{equation}

\subsection{Perturbative potential}

The perturbative potential in momentum space is conventionally expressed as
\begin{equation}
 \tilde V(p,\mu)  =  -C_F\frac{4\pi}{p^2}\,\tilde\alpha_{V}[\alpha_s(\mu),L(\mu,p)]\,, \label{eq:Vppert}
\end{equation}
with
\begin{equation}
 L \equiv  L(\mu,p)  =  \ln\frac{\mu^2}{p^2}\,,
\end{equation}
where $\mu\gg\Lambda_{\rm QCD}$ denotes a a arbitrarily chosen, fixed perturbative momentum scale.
$C_F$ is the eigenvalue of the quadratic Casimir operator for the fundamental representation of the gauge group; $C_F=4/3$ for ${\rm SU}(3)$.
The entire non-trivial structure of $\tilde V(p)$ is encoded in the function $\tilde\alpha_{V}[\alpha_s(\mu),L(\mu,p)]$,
which has the following expansion in powers of $\alpha_s(\mu)$ \cite{Chishtie:2001mf,Anzai:2010td},
\begin{equation}
 \tilde\alpha_{V}[\alpha_s(\mu),L] = \alpha_s(\mu)\sum_{k=0}^{\infty}P_k(L)\left(\frac{\alpha_s(\mu)}{4\pi}\right)^k, \label{eq:alphaV}
\end{equation}
with expansion coefficients $P_k(L)$.

Adapting \Eqref{eq:RG_Vpert} to Eqs.~\eqref{eq:Vppert}-\eqref{eq:alphaV}, we obtain
\begin{equation}
 \left(\frac{\partial}{\partial L}+\frac{\alpha_s}{2}\beta[\alpha_s]\frac{\partial}{\partial \alpha_s}\right)\alpha_s\sum_{k=0}^{\bar k}P_k(L)\left(\frac{\alpha_s}{4\pi}\right)^k={\cal O}(\alpha_s^{\bar k+2}), \label{eq:RGalphaVpert}
\end{equation}
with $\alpha_s\equiv\alpha_s(\mu)$.
This equation constrains the $P_k(L)$ in \Eqref{eq:alphaV} to be polynomials in $L$ of degree $k$, i.e.,
\begin{equation}
 P_k(L)=\sum_{m=0}^{k}\rho_{km}L^m\,, \label{eq:Pk}
\end{equation}
with dimensionless expansion coefficients $\rho_{km}$.
Equation~\eqref{eq:RGalphaVpert} implies that, apart from the explicit values of $a_k\equiv\rho_{k0}$ $(a_0=1)$, the $\rho_{km}$ are fully determined by the coefficients of the $\beta$-function \cite{Chishtie:2001mf}.
For $k\leq3$ they read
\begin{gather}
 \rho_{21}\!=\!(2a_1\beta_0\!+\!\beta_1)\,, \nonumber\\
 \rho_{31}\!=\!(3a_2\beta_0\!+\!2a_1\beta_1\!+\!\beta_2),\ \rho_{32}\!=\!(3a_1\beta_0\!+\!\tfrac{5}{2}\beta_1)\beta_0,
\end{gather}
and $\rho_{kk}=\beta_0^k$.

While for $k\leq2$ the RG equation for $\alpha_s$ is the only source of logarithms encoding a renormalization scale dependence,
for $k\geq3$ the situation is more complex: 
The $P_k(L)$ with $k\geq3$ generically include IR divergences with associated scale $\mu_{\rm us}$ dependent logarithms $L_{\rm US}\equiv L(\mu_{\rm us},p)$.
Thus, besides the logarithms $L$ governed by the RG equation~\eqref{eq:RGalphaVpert}, they depend on extra logarithms $L_{\rm US}$.
We absorb these extra logarithms in the definition of the $a_k$ with $k\geq3$. Particularly for $k=3$, we write $a_3=\bar a_3+a_{3{\rm ln}}L_{\rm US}$ \cite{Anzai:2009tm,Anzai:2010td}.

The coefficients $a_1$ \cite{Fischler:1977yf,Billoire:1979ih} and $a_2$ \cite{Peter:1996ig,Peter:1997me,Schroder:1998vy} are known analytically.
For gauge group $SU(3)$ and in the $\overline{\rm MS}$-scheme, they read 
\begin{align}
 a_1&=\frac{31}{3}-\frac{10}{9}n_f\,, \nonumber\\
 a_2&=\frac{4343}{18}+36\pi^2-\frac{9}{4}\pi^4+66\zeta(3) \nonumber\\
  &\quad-\left(\frac{1229}{27}+\frac{52}{3}\zeta(3)\right)
n_f +\frac{100}{81}n_f^2\,.
\end{align}
 
The coefficients $\bar a_3$ \cite{Smirnov:2008pn,Smirnov:2009fh,Anzai:2009tm,Smirnov:2010zc,Anzai:2010td} and $a_{3{\rm ln}}$ \cite{Appelquist:1977es,Brambilla:1999qa,Brambilla:1999xf} are also known explicitly.
They are, however, not unambiguously fixed, but inherently depend on the scheme used to factorize the US contributions (cf. also the remarks in \cite{Kniehl:2002br,Brambilla:2010pp}).
At order $\alpha_s^{1+k}$, with $k\geq3$, only the sum of the respective contributions of $\cal V$ and ${\cal V}^{\rm US}$, constituting the physical quantity [cf. \Eqref{eq:calE}], is unambiguously determined.
Here we adopt the scheme used by \cite{Brambilla:1999qa,Brambilla:1999xf,Kniehl:1999ud}, where all Fourier transforms are performed in $D=3$ space dimensions and 
the divergent US loop integral (cf. Fig.~\ref{fig:US}) is carried out in $D=3-2\varepsilon$ ($\varepsilon\to0^+$) space dimensions. Specializing to $SU(3)$, the coefficient $\bar a_3$ is then given by \cite{Smirnov:2009fh,Anzai:2009tm}
\begin{equation}
 \bar a_3=a_3^{(0)}+a_3^{(1)}n_f+a_3^{(2)}n_f^2+a_3^{(3)}n_f^3\,,
\end{equation}
with
\begin{align}
a_3^{(0)}&=27c_1 +\frac{15}{16}c_2\,, \nonumber\\
a_3^{(1)}&=\frac{9}{2}c_3+\frac{5}{96}c_4-\frac{68993}{81} +\frac{16624}{27}\zeta(3)+\frac{160}{9}\zeta(5)\,, \nonumber\\
a_3^{(2)}&=\frac{93631}{972}+\frac{16}{45}\pi^4+\frac{412}{9}\zeta(3)\,, \nonumber\\
a_3^{(3)}&=-\frac{1000}{729}\,.
\end{align}
The constants $c_i$ ($i=1\ldots4$) are only known numerically: $c_1$ and $c_2$ have been determined independently by both \cite{Smirnov:2009fh} and \cite{Anzai:2009tm}. The numerical values of \cite{Smirnov:2009fh}, who provide smaller statistical errors, are
\begin{equation}
 c_1 = 502.24(1)\,, \quad
 c_2 = -136.39(12)\,,
\end{equation}
while $c_3$ and $c_4$ read \cite{Smirnov:2008pn},
\begin{equation}
 c_3 = -709.717\,, \quad
 c_4 = -56.83(1)\,.
\end{equation}
Moreover,
\begin{equation}
 a_{3{\rm ln}}=\frac{8}{3}\pi^2C_A^3\,,
\end{equation}
with $C_A=3$ for $SU(3)$.

Thus, at the moment the polynomials~\eqref{eq:Pk} are completely known for $k\leq3$, and therewith the perturbative potential $\tilde V$ up to ${\cal O}(\alpha_s^4)$.

\subsection{Ultrasoft corrections}

The US corrections to the singlet static energy are most conveniently evaluated within the effective field theory framework of potential nonrelativistic QCD (pNRQCD).

Potential NRQCD can be derived rigorously from QCD or more specifically nonrelativistic QCD (NRQCD) \cite{Caswell:1985ui,Bodwin:1994jh}, and provides a formulation of the non-relativistic $Q\bar Q$ system in terms of heavy (static) quark-antiquark composite (color singlet and octet) fields, the associated potentials, light fermions, and US gluons and interactions~\cite{Pineda:1997bj,Brambilla:1999xf,Brambilla:2004jw}.

It is based upon the observation that the $Q\bar Q$ state in the static limit is still characterized by two separated energy scales:
a soft scale of the order of the typical relative momentum of the heavy quarks, and an US scale of the order of the typical binding energy, which is much smaller than the relative momentum.
Hence, even when the soft scale is integrated out to be encoded in perturbative potentials, there is a remnant US sector, which features US gluons and light quarks as dynamical degrees of freedom,
constituting pNRQCD. 
These US degrees of freedom generically induce contributions -- the US corrections -- to the static energy.

In the static limit and up to first order in a multipole expansion in the relative coordinate $\vec{r}$, the pNRQCD Lagrangian for the static $Q\bar Q$ state reads \cite{Pineda:1997bj,Brambilla:1999xf}
\begin{align}
 {\cal L}_{\rm pNRQCD}&={\rm Tr}\Bigl\{S^\dag(i\partial_0-V_s)S + O^\dag(iD_0-V_o)O\Bigr\} \nonumber\\
&\quad+ gV_A {\rm Tr}\Bigl\{O^\dag\vec{r}\cdot\vec{E}S + S^\dag\vec{r}\cdot\vec{E}O \Bigr\} \nonumber\\
&\quad+ g\frac{V_B}{2} {\rm Tr}\Bigl\{O^\dag\vec{r}\cdot\vec{E}O + O^\dag O\vec{r}\cdot\vec{E} \Bigr\} \nonumber\\
&\quad+\sum_l\bar q^l i\slashed{D}q^l - \frac{1}{4}F^a_{\mu\nu}F^{a\mu\nu}. \label{eq:pNRQCD}
\end{align}
Here, the degrees of freedom are static $Q\bar Q$ singlet $S\equiv S(\vec{r},\vec{R},t)$ and octet $O\equiv O(\vec{r},\vec{R},t)$ fields with US energy,
which transform accordingly with respect to gauge transformations in $\vec{R}$, US gluons $A_\mu(\vec{R},t)$ with field strength tensor $F^a_{\mu\nu}\equiv F^a_{\mu\nu}(\vec{R},t)$ and light (massless) quarks of flavor $l$, described by Dirac spinors $q^l$.
Moreover, $\vec{E}^a\equiv\vec{E}^a(\vec{R},t)$, with $a\in\{1,\ldots,8\}$, is the chromoelectric field and $iD_0O\equiv i\partial_0O-g[A_0(\vec{R},t),O]$ the time component of the covariant derivative in the octet representation.
$V_s\equiv V_s(r)$ and $V_o\equiv V_o(r)$ are the singlet and octet static potentials in position space, and $V_A\equiv V_A(r)$, $V_B\equiv V_B(r)$ are dimensionless matching coefficients, to be determined by matching pNRQCD with NRQCD
, i.e., by demanding the physical outcomes of a pNRQCD calculation to agree with those evaluated within NRQCD, e.g., at the same order in perturbation theory. In particular note that $V_s\equiv V$.
For a detailed introduction to pNRQCD we refer the reader to \cite{Brambilla:1999xf,Brambilla:2004jw}.

Decisive in the approach advocated here is the assumption that the perturbative position space potentials in \Eqref{eq:pNRQCD} are defined in terms of a Fourier transform of the respective potentials in momentum space.
Correspondingly, we have
\begin{align}
 V_s&\equiv\int\frac{{\rm d}^3p}{(2\pi)^3}\,{\rm e}^{i\vec{p}\cdot\vec{r}}\left(-C_F\frac{4\pi}{p^2}\alpha_s\right)+{\cal O}(\alpha_s^2), \nonumber\\
 V_o&\equiv\int\!\!\frac{{\rm d}^3p}{(2\pi)^3}\,{\rm e}^{i\vec{p}\cdot\vec{r}}\left[\left(\frac{C_A}{2}-C_F\right)\!\frac{4\pi}{p^2}\alpha_s\right]+{\cal O}(\alpha_s^2).
\end{align}

\begin{figure}[h]
\center
\includegraphics[width=0.28\textwidth]{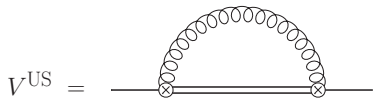} 
\caption{Leading-order US contribution to the singlet static energy evaluated within pNRQCD. The single (double) line represents the singlet (octet) propagator, and the wavy line denotes the US gluon propagator. The relevant singlet-to-octet couplings can be inferred from the Lagrangian~\eqref{eq:pNRQCD}.}
\label{fig:US}
\end{figure}

As suggested by the pNRQCD Lagrangian~\eqref{eq:pNRQCD}, the US corrections are most naturally evaluated in position space.
The Feynman diagram depicted in Fig.~\ref{fig:US} comprises the entire US contribution up to order $\alpha_s^4$, which reads
\begin{multline}
 V^{\rm US}(r,\mu_{\rm us})=-C_F\frac{\alpha_s(\mu)}{\pi}\frac{r^2}{3}V_A^2\left(V_o-V_s\right)^3 \\
\times\left(\ln\frac{(V_o-V_s)^2}{\mu^2_{\rm us}}-\frac{5}{3}+2\ln 2\right). \label{eq:delta_US_form}
\end{multline}

As $V_B$ obviously does not contribute to $V^{\rm US}(r,\mu)$ at the desired order, we will leave it undetermined here. 
The explicit value of $V_A$ follows straightforwardly from a perturbative matching condition. Namely, we demand the singlet static energies calculated in both
NRQCD and pNRQCD to match each other.
In order to fix $V_A$, it suffices to isolate and resum the NRQCD diagrams which give rise to a $\ln(V_o-V_s)$ dependence.  
The corresponding contribution reads \cite{Appelquist:1977es,Brambilla:1999qa}
\begin{equation}
 E^{\rm NRQCD}_{\ln}= -C_FC_A^2\frac{\alpha_s^3}{12\pi}(V_o-V_s)\ln\!\left[(V_o-V_s)^2r^2\right]\!. \label{eq:deltaNRQCD}
\end{equation}

Demanding the prefactors of $\ln(V_o-V_s)$ in Eqs.~\eqref{eq:delta_US_form} and \eqref{eq:deltaNRQCD} to match, we infer
\begin{equation}
 r^2V_A^2(V_o-V_s)^2-\left(\frac{C_A}{2}\alpha_s\right)^2={\cal O}(\alpha_s),
\end{equation}
such that $V_A$ is given by
\begin{equation}
 V_A=\frac{C_A}{2}\frac{\alpha_s}{r}\frac{1}{V_o-V_s} + {\cal O}(\alpha_s), \label{eq:VA}
\end{equation}
and \Eqref{eq:delta_US_form} becomes
\begin{multline}
 V^{\rm US}(r,\mu_{\rm us})=-C_FC_A^2\frac{\alpha_s^3}{12\pi}\left(V_o-V_s\right) \\
\times\left(\ln\frac{(V_o-V_s)^2}{\mu^2_{\rm us}}-\frac{5}{3}+2\ln 2\right). \label{eq:delta_US_form_v2}
\end{multline}
In particular note that this expression features an overall {\it linear} dependence on the potential difference $V_o-V_s$.

This turns out to be absolutely essential in order to guarantee the cancellation of the $\ln\mu^2_{\rm us}$ dependence contained in the coefficient $a_3$ of the perturbative potential with the corresponding term of \Eqref{eq:delta_US_form_v2} (and the associated divergences, somehow obscured in the $\overline{\rm MS}$-scheme)
also when defining the position space potentials via a restricted Fourier transform.

This cancellation ensures the finiteness of the coefficients of $\alpha_s^n$, with $n\leq4$, and $\alpha_s^4\ln(\alpha_s)$ of the singlet static energy in the perturbative regime.
It renders the singlet static energy at ${\cal O}(\alpha_s^4)$ independent of any auxiliary, externally set momentum scale $\mu_{\rm us}$ apart from $\mu_f$.

For completeness, we note that one mighty wonder about the compatibility of \Eqref{eq:VA} with the finding of \cite{Brambilla:1999qa} that $V_A=1+{\cal O}(\alpha_s)$.
We emphasize that \Eqref{eq:VA} is more genuine: When defining the position space potentials in terms of a standard unrestricted Fourier transform it reduces to the expression of \cite{Brambilla:1999qa}.
However, when the position space potential is defined differently, as, e.g., by means of a restricted Fourier transform [cf. \Eqref{eq:VA2} below], or more generally by a given renormalon subtraction scheme, \Eqref{eq:VA} generically deviates from $V_A=1+{\cal O}(\alpha_s)$.
Correspondingly -- and in fact not surprisingly -- also the matching coefficient $V_A$ receives scheme dependent corrections.

\section{Towards the static energy in position space} \label{seq:statEresFT}

All the integrations relevant in performing the restricted Fourier transform from momentum to position space can be traced back to just two generic Fourier integrals, which can be tackled analytically.

To perform the unconstrained Fourier transform, we employ the following identity,
\begin{multline}
 \int\frac{{\rm d}^3p}{(2\pi)^3}\,{\rm e}^{i\vec{p}\cdot\vec{r}}\,\frac{4\pi}{p^2}\,L^m 
=\frac{1}{r}\sum_{j=0}^m\binom{m}{j}\ln^j(r^2\mu^2) \\
\times\!\biggl[\partial_\eta^{m-j}\exp\biggl\{2\gamma_E\eta+\!\sum_{l=2}^\infty\eta^l\tfrac{[2^l-1-(-1)^l]\zeta(l)}{l}\biggr\}\biggr]\biggr|_{\eta=0}, \label{eq:unconstrainedFT}
\end{multline}
which can be derived straightforwardly by combining several identities given in Sec.~6 of \cite{Peter:1997me} and making use of the general Leibniz rule.
Here $\gamma_E$ denotes the Euler-Mascheroni constant and $\zeta(\chi)$ is the Riemann $\zeta$-function.
The notation in \Eqref{eq:unconstrainedFT} is to be understood as follows: First all derivatives for $\eta$ have to be taken. Finally $\eta$ is set to zero. 
In \Eqref{eq:unconstrainedFT} these derivatives can easily be taken explicitly. They amount to multiplications with numerical constants.

As derived in detail in Appendix~\ref{app:restrictedFT}, for $\mu\geq\mu_f$ the other generic Fourier integral can be represented as
\begin{multline}
\int\!\frac{{d^3 p}}{(2\pi)^3}\,{\rm e}^{i\vec{p}\cdot\vec{r}} \,\frac{4\pi}{p^2}\,L^m\,\Theta(\mu_f-|\vec{p}|) \\
=-\frac{\mu_f}{\pi}\sum_{j=0}^{m}\binom{m}{j}\ln^j\frac{\mu^2}{\mu_f^2}\,(-2)^{m-j} \hspace*{2.5cm} \\
\times\left[\partial_\eta^{m-j}\left\{\tfrac{\Gamma(\eta)-\Gamma(\eta,{\rm i}r\mu_f)}{(ir\mu_f)^{1+\eta}}+{\rm c.c.}\right\}\right]\Bigr|_{\eta=0}\,, \label{eq:restrictedFT}
\end{multline}
where $\Gamma(\alpha,\chi)$ denotes the incomplete gamma function. The abbreviation c.c. stands for complex conjugate.
Terms which do not involve any derivatives for $\eta$ can be expressed in a rather compact form,
\begin{equation}
 \left.\left[\frac{\Gamma(\eta)-\Gamma(\eta,ir\mu_f)}{(ir\mu_f)^{1+\eta}}+{\rm c.c.}\right]\right|_{\eta=0}=-2\,\frac{{\rm Si}(r\mu_f)}{r\mu_f}, \label{eq:Si}
\end{equation}
where ${\rm Si}(\chi)=\int_0^\chi{\rm d}t\,\frac{\sin t}{t}$ is the sine integral.
Contributions with derivatives acting do not have particularly nice explicit representations. They can be written in terms of hypergeometric functions (cf. Appendix~\ref{app:seriesrep}).
For them we prefer to keep the above representation involving parameter differentiations. For series representations, see Appendix~\ref{app:seriesrep}.

We emphasize that the condition $\mu\geq\mu_f$ invoked in the derivation of \Eqref{eq:restrictedFT} is fully compatible with our objectives:
As $\mu$ corresponds to a momentum scale in the perturbative regime, that in a sense can be seen as bounded from below by $\mu_f$ [cf. the discussion in the context of Eqs.~\eqref{eq:V(r)}-\eqref{eq:deltaV}],
the ratio $\mu/\mu_f$ is generically equal or larger than one. Hence, we will implicitly assume this condition to be fulfilled in the remainder.

Note that the structure of Eqs.~\eqref{eq:unconstrainedFT} and \eqref{eq:restrictedFT} is quite similar, in the sense that 
the entire $\mu$ dependence is encoded in logarithms of $\mu^2$, rendered dimensionless by an additional momentum-scale squared. Both expressions amount to series expansions in powers of these logarithms.

\subsection{Perturbative potential in position space}

With the help of Eqs.~\eqref{eq:unconstrainedFT} and \eqref{eq:restrictedFT} it is straightforward to explicitly determine both $V(r)$ and $\delta V(r,\mu_f)$ at the accuracy with which the perturbative potential in momentum space is known,  i.e., presently up to order $\alpha_s^4$. The perturbative potential in position space defined by a restricted Fourier transform~\eqref{eq:resFT} is obtained by subtracting these results.

In order to allow for a clear and compact representation of the results, we introduce the following shortcut notation for the $n$th derivative of the polynomials $P_k(L)$ from \Eqref{eq:Pk}:  $P_k^{(n)}(L)=\frac{\partial^n}{\partial L^n} P_k(L)$.
The $P_k^{(n)}(L)$ with $n=k$ are independent of $L$ and read $P_k^{(k)}\equiv P_k^{(k)}(L)=k!\beta_0^k$.

Hence, $V(r)$ and $\delta V(r,\mu_f)$ are known with the following accuracy,
\begin{widetext}
\begin{multline}
V(r)=-C_F\frac{\alpha_s(\mu)}{r}
\Biggl\{
1+\frac{\alpha_s(\mu)}{4\pi}\biggl[P_1\Bigl(\ln(r^2\mu^2)\Bigr)+P_1'\partial_\eta\biggr] \\
\quad\!+\left(\frac{\alpha_s(\mu)}{4\pi}\right)^2\biggl[P_2\Bigl(\ln(r^2\mu^2)\Bigr)
+P_2'\Bigl(\ln(r^2\mu^2)\Bigr)\partial_\eta +\frac{1}{2}P_2''\partial_\eta^2\biggr] \\
+\left(\frac{\alpha_s(\mu)}{4\pi}\right)^3\!\biggl[a_{3{\rm ln}}\Bigl(\ln(r^2\mu_{\rm us}^2)+\partial_\eta\Bigr)+\bar P_3\Bigl(\ln(r^2\mu^2)\Bigr)+P_3'\Bigl(\ln(r^2\mu^2)\Bigr)\partial_\eta
  +\frac{1}{2}P_3''\Bigl(\ln(r^2\mu^2)\Bigr)\partial_\eta^2 +\frac{1}{6}P_3'''\partial_\eta^3\biggr]
\Biggr\}\!\! \\
\times\exp\biggl\{2\gamma_E\eta+\!\sum_{l=2}^\infty\eta^l\tfrac{[2^l-1-(-1)^l]\zeta(l)}{l}\biggr\}\biggr|_{\eta=0}, \label{eq:Vpert_expl}
\end{multline}
\vspace*{1mm}
and
\vspace*{1mm}
\begin{multline}
\delta V(r,\mu_f)=C_F\frac{\alpha_s(\mu)}{\pi}\mu_f
\Biggl\{
1+\frac{\alpha_s(\mu)}{4\pi}\biggl[P_1\Bigl(\ln\tfrac{\mu^2}{\mu_f^2}\Bigr)-2P_1'\partial_\eta\biggr] \\
\quad\!+\left(\frac{\alpha_s(\mu)}{4\pi}\right)^2\biggl[P_2\Bigl(\ln\tfrac{\mu^2}{\mu_f^2}\Bigr)
-2P_2'\Bigl(\ln\tfrac{\mu^2}{\mu_f^2}\Bigr)\partial_\eta +2P_2''\partial_\eta^2\biggr] \\
+\left(\frac{\alpha_s(\mu)}{4\pi}\right)^3\!\biggl[a_{3{\rm ln}}\Bigl(\ln\tfrac{\mu^2_{\rm us}}{\mu^2_f}+\partial_\eta\Bigr)+\bar P_3\Bigl(\ln\tfrac{\mu^2}{\mu_f^2}\Bigr)-2P_3'\Bigl(\ln\tfrac{\mu^2}{\mu_f^2}\Bigr)\partial_\eta
  +2P_3''\Bigl(\ln\tfrac{\mu^2}{\mu_f^2}\Bigr)\partial_\eta^2 -\frac{4}{3}P_3'''\partial_\eta^3\biggr]
\Biggr\} \\
\times\left.\left[\frac{\Gamma(\eta)-\Gamma(\eta,ir\mu_f)}{(ir\mu_f)^{1+\eta}}+{\rm c.c.}\right]\right|_{\eta=0}, \label{eq:step5}
\end{multline}
\vspace*{2mm}
\end{widetext}
where we explicitly separated the $\mu_{\rm us}$ dependence at ${\cal O}(\alpha_s^4)$ off the  polynomial $P_3(L)$, by introducing
$\bar P_3(L)=P_3(L)-a_{3{\rm ln}}L_{\rm US}$. Of course, $P^{(n)}_3(L)=\bar P^{(n)}_3(L)$ for $n\geq1$.

Let us briefly comment on the general structure of $V$ and $\delta V$ represented in the fashion of Eqs.~\eqref{eq:Vpert_expl} and \eqref{eq:step5}:
Polynomials $P_k^{(n)}(L)$ come along with $n$th derivatives $\partial_{\eta}^n$ of the respective master integral expressions from Eqs.~\eqref{eq:unconstrainedFT} and \eqref{eq:restrictedFT}.
The contribution at a given, fixed order in the coupling $\sim\alpha_s^{1+k}$, with $k\geq0$, amounts to a sum of terms $\sim P_m^{(n)}(L)\partial_\eta^m$ with $0\leq m\leq k$ acting on the master integral expressions.
Thus, the highest derivative at order $\alpha_s^{1+k}$ is a $k$th derivative contribution $\sim k!\beta_0^k\partial_\eta^k$. This factorial growth has also been noted by \cite{Beneke:1998rk}.

So far we only focused on the formal structure of the subtraction term $\delta V(r,\mu_f)$.
In order to allow for some more quantitative insights into $\delta V(r,\mu_f)$ up to order $\alpha_s^4$, we exemplarily set $n_f=2$, $\mu=\mu_f$, $\alpha_s\equiv\alpha_s(\mu_f)$ and resort to an expansion in powers of $r^2\mu_f^2$.
The particular choice $\mu=\mu_f$ implies $\ln\frac{\mu^2}{\mu_f^2}=0$, which allows us to render the prefactors of powers of $\alpha_s$ in a particularly simple form and to give most of them numerically,
\begin{multline}
\delta V(r,\mu_f)\approx-2\mu_f\frac{C_F}{\pi}\alpha_s \\
\times\biggl\{
1+2.18\,\alpha_s+9.77\,\alpha_s^2 + \Bigl(53.72+0.36\,\ln\tfrac{\mu_{\rm us}^2}{\mu_f^2}\Bigr)\alpha_s^3 \\
-\frac{r^2\mu_f^2}{18}\biggl[1+1.16\,\alpha_s+3.59\,\alpha_s^2 \\ +\Bigl(10.28+0.36\,\ln\tfrac{\mu_{\rm us}^2}{\mu_f^2}\Bigr)\alpha_s^3
\biggr]
+{\cal O}(r^4\mu_f^4) \biggr\}. \label{eq:deltaVexpl}
\end{multline}
Obviously, the coefficients of $\alpha_s^{1+k}$ with $k\in\mathbb{N}$ increase with $k$.
Even though this increase is more pronounced for the contribution $\sim\mu_f$, it is also clearly visible for the contribution $\sim\mu_f(r^2\mu_f^2)$.
Given the present accuracy of $\tilde V(p)$, we can of course only provide explicit values for the coefficients up to ${\cal O}(\alpha_s^4)$.
However, we can at least study the behavior of the expansions coefficients $d_l(n)$ of the expression
\begin{multline}
\biggl[\partial_\eta^n\biggl\{\frac{\Gamma(\eta)- \Gamma(\eta,ir\mu_f)}{(ir\mu_f)^{1+\eta}}+{\rm c.c.}\biggr\}\biggr]\bigg|_{\eta=0} \\
=\sum_{l=0}^\infty d_l(n)(r^2\mu_f^2)^{l} \label{eq:diffm_masterintexp1_maintext}
\end{multline}
in \Eqref{eq:step5}. The derivation of \Eqref{eq:diffm_masterintexp1_maintext} is given in Appendix~\ref{app:seriesrep}. Noteworthily, the coefficients $d_l(n)$ are explicitly known for any values of $\{n,l\}\in\mathbb{N}_0$.
As detailed in Appendix~\ref{app:closerlook}, the modulus of the coefficients $d_0(n)$ is given by $|d_0(n)|=2n!$, such that the coefficients for $l=0$ increase factorial with $n$. Moreover, for a given $l\geq1$ the coefficients $d_l(n)$ fulfill $|d_{l+1}(n)|>|d_l(n)|$ if $n>2l$.

As noted in the context of \Eqref{eq:step5}, at order $\alpha_s^{1+k}$ in $\delta V(r,\mu_f)$ the highest derivative for $\eta$ is $\partial_\eta^k$ with prefactor $\sim k!\beta_0^k$.
Thus, the modulus of the contribution $\sim(r^2\mu_f^2)^0$ in
\begin{equation}
 P^{(k)}_k\biggl[\partial_\eta^k\biggl\{\frac{\Gamma(\eta)- \Gamma(\eta,ir\mu_f)}{(ir\mu_f)^{1+\eta}}+{\rm c.c.}\biggr\}\biggr]\bigg|_{\eta=0}
\end{equation}
increases by a factor of $(k+1)^2\beta_0$ when shifting $k\to k+1$, while the moduli of the contributions $\sim(r^2\mu_f^2)^l$ increase by a somewhat smaller factor of $\frac{1}{2l+1}(k+1)^2\beta_0$ for $k>2l$  (cf. Appendix~\ref{app:closerlook}).
For large enough values of $k$ we expect this behavior to let the moduli of the expansion coefficients of a given contribution $\sim \mu_f(r^2\mu_f^2)^l\alpha_s^{1+k}$ to $\delta V(r,\mu_f)$,
and correspondingly the uncontrolled contributions $\sim\Lambda_{\rm QCD}(r^2\Lambda_{\rm QCD}^2)^l\alpha_s^{1+k}$ contained in the potential $V(r)$ as defined by an unconstrained Fourier transform,
grow when increasing $k\to k+1$, while keeping $l$ fixed.
We consider this as an additional, conceptual motivation to define the potential in position space by means of a restricted Fourier transform, allowing for a complete subtraction of these potentially pathologically behaving terms originating in the uncontrolled low momentum contributions of the Fourier integral~\eqref{eq:deltaV}, particularly also at higher orders.

In the next step we want to show that the expressions~\eqref{eq:Vpert_expl} and \eqref{eq:step5} form RG invariants with respect to the renormalization scale $\mu$, i.e., fulfill
\begin{equation}
 \mu\frac{d}{d \mu}V(r)={\cal O}(\alpha^5)\,, \quad \mu\frac{d}{d \mu}\delta V(r,\mu_f)={\cal O}(\alpha^5)\,. \label{eq:mu/dmuVdeltaV}
\end{equation}
This basically follows from \Eqref{eq:RGalphaVpert}, which implies
\begin{multline}
 \mu\frac{d}{d \mu}\Biggl\{\alpha_s(\mu)\Biggl[\sum_{k=0}^2 P_k(L)\left(\frac{\alpha_s(\mu)}{4\pi}\right)^k \\
+ \bar P_3(L)\left(\frac{\alpha_s(\mu)}{4\pi}\right)^3\Biggr]\Biggr\}={\cal O}(\alpha_s^5)\,. \label{eq:RGinv}
\end{multline}
As the entire $p^2$ dependence of \Eqref{eq:RGinv} is via $L$, we moreover have
\begin{equation}
 -p^2\frac{d}{dp^2}=\frac{d}{dL}\,. \label{eq:d/dL}
\end{equation}
Applying $n\in\{1,2,3\}$ times the derivative operator~\eqref{eq:d/dL} onto \Eqref{eq:RGinv}, it is straightforward to show that also 
\begin{equation}
  \mu\frac{d}{d \mu}\Biggl\{\alpha_s(\mu)\Biggl[\sum_{k=n}^3 P_k^{(n)}(L)\left(\frac{\alpha_s(\mu)}{4\pi}\right)^k \Biggr]\Biggr\}={\cal O}(\alpha_s^5) \label{eq:derivRGinv}
\end{equation}
holds [cf. \Eqref{eq:Pk}].
Employing Eqs.~\eqref{eq:RGinv} and \eqref{eq:derivRGinv} in Eqs.~\eqref{eq:Vpert_expl} and \eqref{eq:step5} we obtain \Eqref{eq:mu/dmuVdeltaV}.

Let us however emphasize again that an essential feature of Eqs.~\eqref{eq:Vpert_expl} and \eqref{eq:step5} is their dependence on the additional renormalization scale $\mu_{\rm us}$, which ultimately is to be canceled in the expression of the singlet static energy.

For completeness, we note also that the result for $\delta V(\mu_f)$ as derived by \cite{Beneke:1998rk} can be reproduced most conveniently by replacing
\begin{equation}
 \frac{\Gamma(\eta)-\Gamma(\eta,ir\mu_f)}{(ir\mu_f)^{1+\eta}}+{\rm c.c.}\ \to\ -\frac{2}{1+\eta} \label{eq:backtodeltaV(muf)}
\end{equation}
in \Eqref{eq:step5}, i.e., by substituting the expression on the left-hand side of \Eqref{eq:backtodeltaV(muf)} by its value at $r=0$,
such that $\delta V(\mu_f)=\delta V(r=0,\mu_f)$ (cf. Appendix~\ref{app:seriesrep}).

\subsection{Ultrasoft corrections in the restricted Fourier transform scheme}

The US correction to the static energy at ${\cal O}(\alpha_s^4)$ is straightforwardly obtained form \Eqref{eq:delta_US_form_v2}.
In order to adopt it to the restricted Fourier transform scheme advocated here, we simply have to substitute the potential difference $V_o-V_s$ by the leading order result of an restricted Fourier transform~\eqref{eq:resFT},
\begin{equation}
 (V_o- V_s)(r,\mu_f)=\frac{C_A}{2}\frac{\alpha_s}{r}\!\left(1-\frac{2}{\pi}{\rm Si}(r\mu_f)\right)+{\cal O}(\alpha_s^2)\,,
\end{equation}
whereform we also infer [cf.~\Eqref{eq:VA}]
\begin{equation}
 V_A(r,\mu_f)=\frac{1}{1-\frac{2}{\pi}\,{\rm Si}(r\mu_f)}+{\cal O}(\alpha_s). \label{eq:VA2}
\end{equation}
For \Eqref{eq:delta_US_form_expl} this implies
\begin{multline}
 V^{\rm US}(r,\mu_f)=C_F\frac{\alpha_s}{r}\left(\frac{\alpha_s}{4\pi}\right)^3a_{3{\rm ln}}\left(1-\frac{2}{\pi}\,{\rm Si}(r\mu_f)\right) \\
\times\biggl[\ln(r^2\mu^2_{\rm us})-2\ln\left(C_A\alpha_s\right) \\ 
+\frac{5}{3}-2\ln\left(1-\frac{2}{\pi}\,{\rm Si}(r\mu_f)\right)\biggr]. \label{eq:delta_US_form_expl}
\end{multline}
In the limit $\mu_f\to0$ the sine integral vanishes, ${\rm Si}(0)=0$, and we reobtain the expression for the US contribution as, e.g., given explicitly in Eq.~(34) of \cite{Brambilla:2009bi}, which we reproduce (in a slightly different representation) in \Eqref{eq:delta_US_form_expl_1} below.
In analogy to the perturbative potential, we split the US contribution~\eqref{eq:delta_US_form_expl} into the standard expression as obtained by an unconstrained Fourier transform,
\begin{multline}
 V^{\rm US}(r)=C_F\frac{\alpha_s}{r}\left(\frac{\alpha_s}{4\pi}\right)^3a_{3{\rm ln}} \\
\times\biggl[\ln(r^2\mu^2_{\rm us})-2\ln\left(C_A\alpha_s\right)+\frac{5}{3}\biggr], \label{eq:delta_US_form_expl_1}
\end{multline}
and a correction $\delta V^{\rm US}(r,\mu_f)$ encoding the entire $\mu_f$ dependence,
defined as follows [cf. \Eqref{eq:resFT}],
\begin{equation}
 \delta V^{\rm US}(r,\mu_f)\equiv V^{\rm US}(r)-V^{\rm US}(r,\mu_f). \label{eq:deltaVUS}
\end{equation}
With foresight we express it in the following rather complicated form
\begin{multline}
 \delta V^{\rm US}(r,\mu_f)=C_F\frac{\alpha_s}{r}\left(\frac{\alpha_s}{4\pi}\right)^3a_{3{\rm ln}} \\
\times
2\left(1-\frac{2}{\pi}{\rm Si}(r\mu_f)\right)\ln\left(1-\frac{2}{\pi}\,{\rm Si}(r\mu_f)\right)\\
-C_F\frac{\alpha_s}{\pi}\mu_f\left(\frac{\alpha_s}{4\pi}\right)^3a_{3{\rm ln}}\biggl[\ln(r^2\mu^2_{\rm us})-2\ln\left(C_A\alpha_s\right)+\frac{5}{3}\biggr]\\
\times\left.\left[\frac{\Gamma(\eta)-\Gamma(\eta,ir\mu_f)}{(ir\mu_f)^{1+\eta}}+{\rm c.c.}\right]\right|_{\eta=0}, \label{eq:deltaV_US}
\end{multline}
where we made use of \Eqref{eq:Si}.

As Eqs.~\eqref{eq:delta_US_form_expl_1} and \eqref{eq:deltaV_US} both feature an overall factor of $\alpha_s^4$, they are clearly invariant of the explicit momentum scale, 
at which the coupling $\alpha_s$ is evaluated up to ${\cal O}(\alpha_s^4)$. Thus, only their dependence on the scale $\mu_{\rm us}$ is of further interest here. 

For completeness, we also provide the expression for $\delta V^{\rm US}(r,\mu_f)$ at order $\alpha_s^4$ in the fashion of \Eqref{eq:deltaVexpl}, i.e.,  we exemplarily set $n_f=2$, and resort to an expansion in powers of $r\mu_f$.
For \Eqref{eq:deltaV_US} this results in
\begin{multline}
\delta V^{\rm US}(r,\mu_f)\approx-2\mu_f\frac{C_F}{\pi}\alpha_s^4 \\
\times\biggr\{
0.91-0.36\ln(r^2\mu_{\rm us}^2)+0.72\ln\alpha_s
- 0.23\,r\mu_f \\
-\frac{r^2\mu_f^2}{18}\Bigl[1.78-0.36\ln(r^2\mu_{\rm us}^2) \\+0.72\ln\alpha_s\Bigr]
+{\cal O}(r^2\mu_f^3) \biggr\}. \label{eq:deltaVUSexpl}
\end{multline}
Note that, contrarily to $\delta V(r,\mu_f)$, the expansion of $\delta V^{\rm US}(r,\mu_f)$ is not only in even powers of $r\mu_f$. Instead also odd powers of $r\mu_f$ show up.

\subsection{The static energy in position space}

The results of the previous sections imply that the perturbative singlet $\bar QQ$ static energy in position space, as defined by a restricted Fourier transform, can conveniently be written as [cf. \Eqref{eq:resFT-E}]
\begin{equation}
 E(r,\mu_f)=E(r)-\delta E(r,\mu_f)\,, \label{eq:pertEres}
\end{equation}
where $E(r)=V(r)-V^{\rm US}(r)$ is the standard result for the singlet static energy as obtained from an unrestricted Fourier transform [cf. \Eqref{eq:E_sFT}].
The correction $\delta E(r,\mu_f)=\delta V(r,\mu_f)-\delta V^{\rm US}(r,\mu_f)$ encodes the differences between $E(r)$ and the static energy as defined by means of a restricted Fourier transform; cf. Eqs.~\eqref{eq:resFT} and \eqref{eq:deltaVUS}.
Combining \Eqref{eq:Vpert_expl} with \Eqref{eq:delta_US_form_expl_1}, and \Eqref{eq:step5} with \Eqref{eq:deltaV_US}, we straightforwardly obtain the explicit results for $E(r,\mu_f)$ and $\delta E(r,\mu_f)$.
They read,
\begin{widetext}
\begin{multline}
E(r)=-C_F\frac{\alpha_s(\mu)}{r}
\Biggl\{
1+\frac{\alpha_s(\mu)}{4\pi}\biggl[P_1\Bigl(\ln(r^2\mu^2)\Bigr)+P_1'\partial_\eta\biggr] \\
\quad\!+\left(\frac{\alpha_s(\mu)}{4\pi}\right)^2\biggl[P_2\Bigl(\ln(r^2\mu^2)\Bigr)
+P_2'\Bigl(\ln(r^2\mu^2)\Bigr)\partial_\eta +\frac{1}{2}P_2''\partial_\eta^2\biggr] \\
+\left(\frac{\alpha_s(\mu)}{4\pi}\right)^3\!\biggl[a_{3{\rm ln}}\Bigl(2\ln\left(C_A\alpha_s\right)-\frac{5}{3}+\partial_\eta\Bigr)+\bar P_3\Bigl(\ln(r^2\mu^2)\Bigr)+P_3'\Bigl(\ln(r^2\mu^2)\Bigr)\partial_\eta
  +\frac{1}{2}P_3''\Bigl(\ln(r^2\mu^2)\Bigr)\partial_\eta^2 +\frac{1}{6}P_3'''\partial_\eta^3\biggr]
\Biggr\}\!\! \\
\times\exp\biggl\{2\gamma_E\eta+\!\sum_{l=2}^\infty\eta^l\tfrac{[2^l-1-(-1)^l]\zeta(l)}{l}\biggr\}\biggr|_{\eta=0}, \label{eq:Eunconstrained_expl}
\end{multline}
and
\begin{multline}
\delta E(r,\mu_f)=C_F\frac{\alpha_s(\mu)}{\pi}\mu_f
\Biggl\{
1+\frac{\alpha_s(\mu)}{4\pi}\biggl[P_1\Bigl(\ln\tfrac{\mu^2}{\mu_f^2}\Bigr)-2P_1'\partial_\eta\biggr] \\
\quad\!+\left(\frac{\alpha_s(\mu)}{4\pi}\right)^2\biggl[P_2\Bigl(\ln\tfrac{\mu^2}{\mu_f^2}\Bigr)
-2P_2'\Bigl(\ln\tfrac{\mu^2}{\mu_f^2}\Bigr)\partial_\eta +2P_2''\partial_\eta^2\biggr] \\
+\left(\frac{\alpha_s(\mu)}{4\pi}\right)^3\!\biggl[a_{3{\rm ln}}\Bigl(2\ln\left(C_A\alpha_s\right)-\ln(r^2\mu_f^2)-\frac{5}{3}+\partial_\eta\Bigr) +\bar P_3\Bigl(\ln\tfrac{\mu^2}{\mu_f^2}\Bigr)-2P_3'\Bigl(\ln\tfrac{\mu^2}{\mu_f^2}\Bigr)\partial_\eta
  +2P_3''\Bigl(\ln\tfrac{\mu^2}{\mu_f^2}\Bigr)\partial_\eta^2 -\frac{4}{3}P_3'''\partial_\eta^3\biggr]
\Biggr\} \\
\times\left.\left[\frac{\Gamma(\eta)-\Gamma(\eta,ir\mu_f)}{(ir\mu_f)^{1+\eta}}+{\rm c.c.}\right]\right|_{\eta=0}
+C_F\frac{\alpha_s(\mu)}{r}\left(\frac{\alpha_s(\mu)}{4\pi}\right)^3 2a_{3{\rm ln}}\left(1-\frac{2}{\pi}{\rm Si}(r\mu_f)\right)\ln\left(1-\frac{2}{\pi}\,{\rm Si}(r\mu_f)\right). \label{eq:deltaE_expl}
\end{multline}
\end{widetext}
Obviously, the $\mu_{\rm us}$ dependences of the respective perturbative and US contributions at ${\cal O}(\alpha_s^4)$ exactly cancel, such that both expressions separately are manifestly rendered independent of the RG scale $\mu_{\rm us}$ also.
Thus, apart from the dependence on the momentum cut-off scale $\mu_f$ inherent to our approach, Eqs.~\eqref{eq:Eunconstrained_expl} and \eqref{eq:deltaE_expl} constitute an overall RG scale independent singlet static energy $E(r,\mu_f)$ in position space up to ${\cal O}(\alpha_s^4)$.

\section{Results and Discussion} \label{seq:Ex+Res}

Having formally derived the $Q\bar Q$ static energy in position space by means of a restricted Fourier transform, we now aim at discussing some exemplary results.
In order to do this, we first need an explicit expression of the coupling $\alpha_s(\mu)$ evaluated at the momentum $\mu$. 

Integrating \Eqref{eq:beta} and iteratively solving it with the integration constant adopted to the $\overline{\rm MS}$ scheme \cite{Bardeen:1978yd,Furmanski:1981cw}, one obtains \cite{Chetyrkin:1997sg}
\begin{multline}
 \alpha_s(\mu)=\frac{4\pi}{\beta_0 l}\biggl\{1-\frac{\beta_1}{\beta_0^2 l}\ln l+\Bigl(\frac{\beta_1}{\beta_0^2l}\Bigr)^2\biggl[\ln^2l-\ln l -1+\frac{\beta_0\beta_2}{\beta_1^2}\biggr]\\
\hfill-\Bigl(\frac{\beta_1}{\beta_0^2l}\Bigr)^3\biggl[\ln^3l-\frac{5}{2}\ln^2l
-\Bigl(2-\frac{3\beta_0\beta_2}{\beta_1^2}\Bigr)\!\ln l \\
+\frac{1}{2}\Bigl(1-\frac{\beta_0^2\beta_3}{\beta_1^3}\Bigr)\biggr] +{\cal O}\Bigl(\frac{1}{l^4}\Bigr)\biggr\}, \label{eq:alpha_s}
\end{multline}
with $l\equiv\ln(\mu^2/\Lambda_{\overline{\rm MS}}^2)$.

Subsequently, we will employ this equation to determine $\alpha_s(\mu)$ at a given perturbative momentum scale $\mu$.
To keep the expressions as compact and concise as possible, we limit our discussion to a fixed number of light flavors; -- somewhat arbitrarily -- we choose to work with $n_f=2$ and set $\Lambda_{\overline{\rm MS}}=315\,{\rm MeV}$ \cite{Jansen:2011vv}.

In order to prevent the logarithms $\sim\ln^m(r^2\mu^2)$ in \Eqref{eq:Eunconstrained_expl} [cf. \Eqref{eq:Pk}] to become large and spoil the perturbative expansion, in the remainder of this paper we moreover identify $\mu=\frac{1}{r}$,
which in turn implies $\ln(r^2\mu^2)\to 0$. The condition $\mu\geq\mu_f$ invoked in the construction of the restricted Fourier transform in Sec.~\ref{seq:statEresFT} thus translates into $r\leq1/\mu_f$.

Correspondingly, the only remaining parameter to be adjusted is the cut-off momentum $\mu_f$.
Aiming at phenomenological applications it seems reasonable to fix $\mu_f$ by resorting to a lattice simulation of the singlet static energy in position space with the given number $n_f$ of dynamical light flavors:
Assuming the perturbative static energy~\eqref{eq:pertEres} as defined by a restricted Fourier transform to reliably describe the {\it true} static energy for (at least) $r\leq1/\mu_f$, and analogously the lattice data to correspond to the true static energy for $r\geq1/\mu_f$, $\mu_f$ could, e.g., be fixed by requiring the first derivative of the static energy in position space to be continuous at the matching point $r=1/\mu_f$.

Such a comparison with lattice data is outside the scope of this paper, which aims at working out the full analytic expression of the perturbative static energy in position space as defined by a restricted Fourier transform at the accuracy with which the perturbative static energy in momentum space is known.

For completeness let us remark that exactly this idea has been pursued and successfully implemented by \cite{Laschka:2011zr} to construct the {\it full potential} in position space for $n_f=0$, by merging perturbation theory with lattice data from \cite{Bali:2000vr}.
However, there are some important differences (which should not distort the possibility of such an approach): Reference~\cite{Laschka:2011zr} performed the restricted Fourier transform only numerically and for $n_f=0$.
Contrarily our analytical results can straightforwardly be adapted to any given number $n_f$ of light flavors.
Moreover, Ref.~\cite{Laschka:2011zr} do not account for the US contributions ${\cal V}^{\rm US}$ -- whose consistent inclusion in the restricted Fourier transform scheme is a major advance of our work -- but only limit themselves to the strictly perturbative contributions ${\cal V}$.

\begin{figure}[h]
\center
\includegraphics[width=0.5\textwidth]{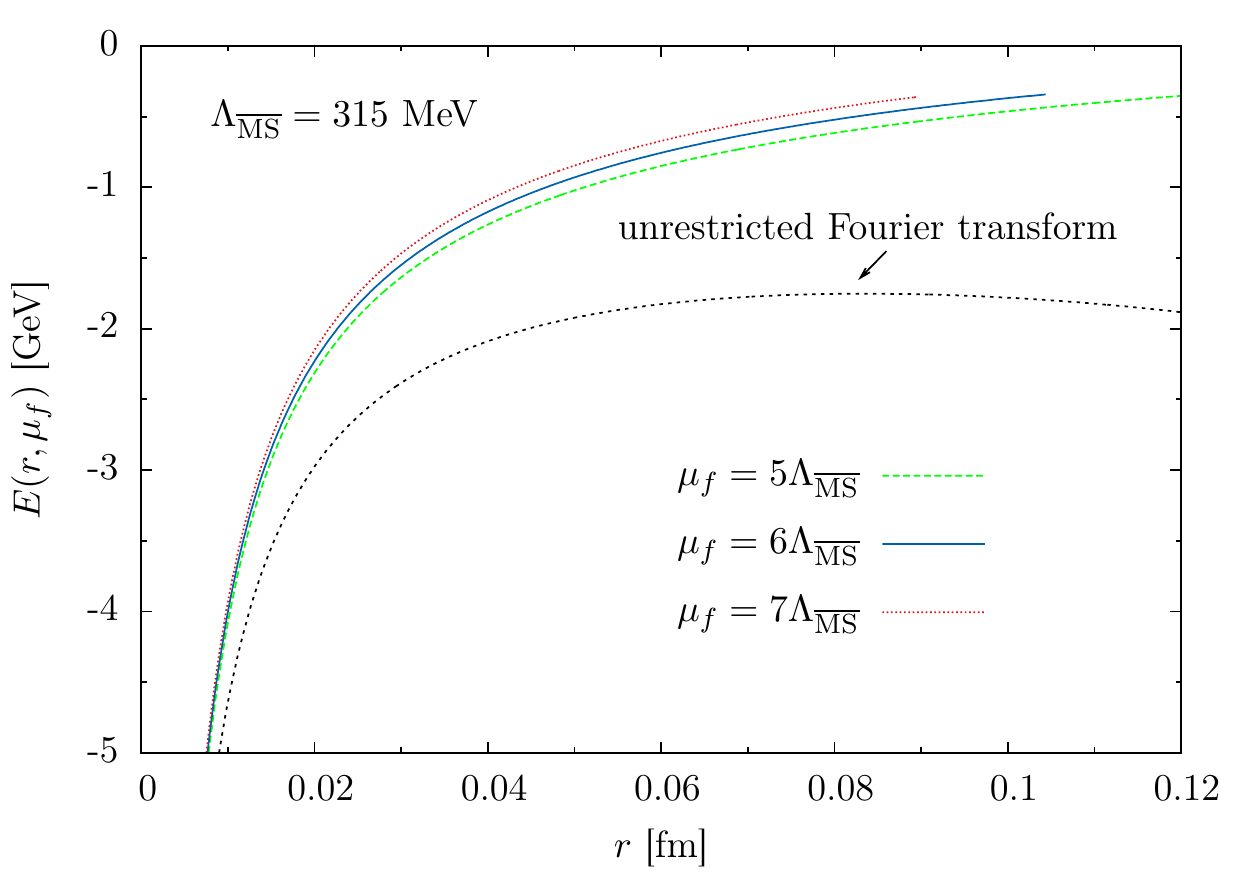} 
\caption{The singlet static energy as obtained from a restricted and a unrestricted Fourier transform from momentum to position space. Obviously, the change of the results with respect to variations of the cut-off momentum scale $\mu_f$ is rather mild
-- all curves with $\mu_f\in[5\Lambda_{\overline{\rm MS}}\ldots7\Lambda_{\overline{\rm MS}}]\approx[1.6\ldots2.2]{\rm GeV}$ fall in the region delimited by the red (upper) dotted and green dashed curves.
For a given value of $\mu_f$, the corresponding curve is depicted for $r\leq1/\mu_f$.}
\label{fig:E}
\end{figure}

In order to give a visual impression of the perturbative static energy in position space, in Fig.~\ref{fig:E} we set $\mu_f=\{5,6,7\}\Lambda_{\overline{\rm MS}}\approx\{1.6,1.9,2.2\}{\rm GeV}$. Note that the couplings evaluated at this scale read $\alpha_s(\mu_f)\approx\{0.30,0.27,0.25\}$. For comparison we also depict the result of an unrestricted Fourier transform~\eqref{eq:Eunconstrained_expl}, which reaches a maximum and starts bending back at $r\approx0.08\,{\rm fm}$.
Conversely, the curves obtained by restricted Fourier transforms basically fall on top of each other for $r\lesssim0.01\,{\rm fm}$, and show the same qualitative behavior for $r\gtrsim0.01\,{\rm fm}$.

\begin{figure}[h]
\center
\includegraphics[width=0.5\textwidth]{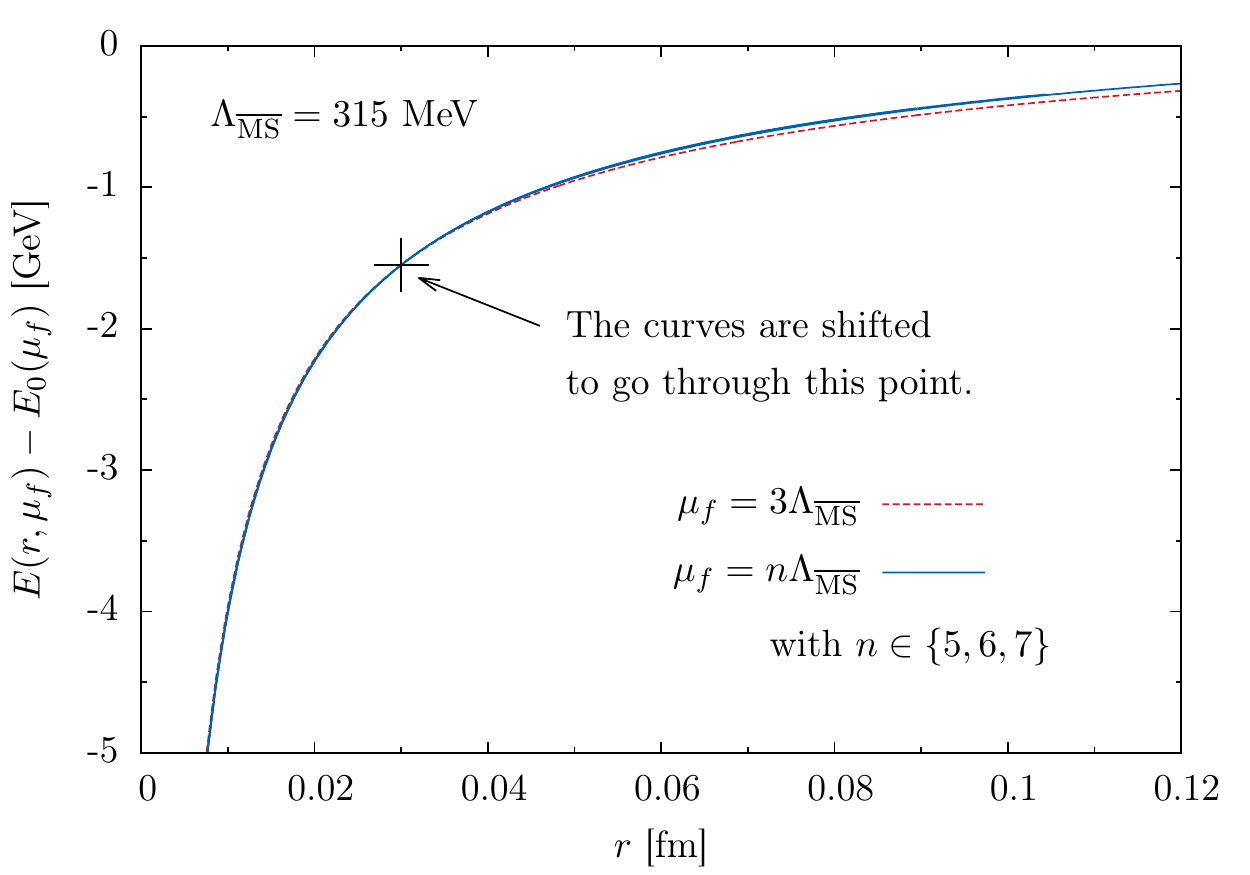} 
\caption{The singlet static energy for various choices of $\mu_f$. The curves for different values of $\mu_f$ are shifted to go through the point marked with the cross by subtracting suitable, $r$ independent constants $E_0(\mu_f)$,
determined as $E_0(\mu_f)=E(r\approx0.03{\rm fm},\mu_f)-E(r\approx0.03{\rm fm},\mu_f=6\Lambda_{\overline{\rm MS}})$. The results for $\mu_f=\{5,6,7\}\Lambda_{\overline{\rm MS}}$ basically fall on top of each other and cannot be discerned in this figure.} 
\label{fig:E_shifted}
\end{figure}

The similarity of the curves for different values of $\mu_f$ becomes even more obvious if we allow for $r$ independent, and thus not measurable, overall energy shifts of our results, as, e.g., necessary when comparing with lattice data.
In Fig.~\ref{fig:E_shifted} we somewhat arbitrarily demanded the various curves to agree at $r\approx0.03\,{\rm fm}$. The differences in the results for $\mu_f=\{5,6,7\}\Lambda_{\overline{\rm MS}}$ in Fig.~\ref{fig:E_shifted} then become practically indiscernible by eye.
Only the curve for $\mu_f=3\Lambda_{\overline{\rm MS}}\approx0.9\,{\rm GeV}$ (note that for this scale the coupling is already as large as $\alpha_s(3\Lambda_{\overline{\rm MS}})\approx0.46$) starts to slightly deviate from the other curves for $r\gtrsim0.05\,{\rm fm}$.

\begin{figure}[h]
\center
\includegraphics[width=0.5\textwidth]{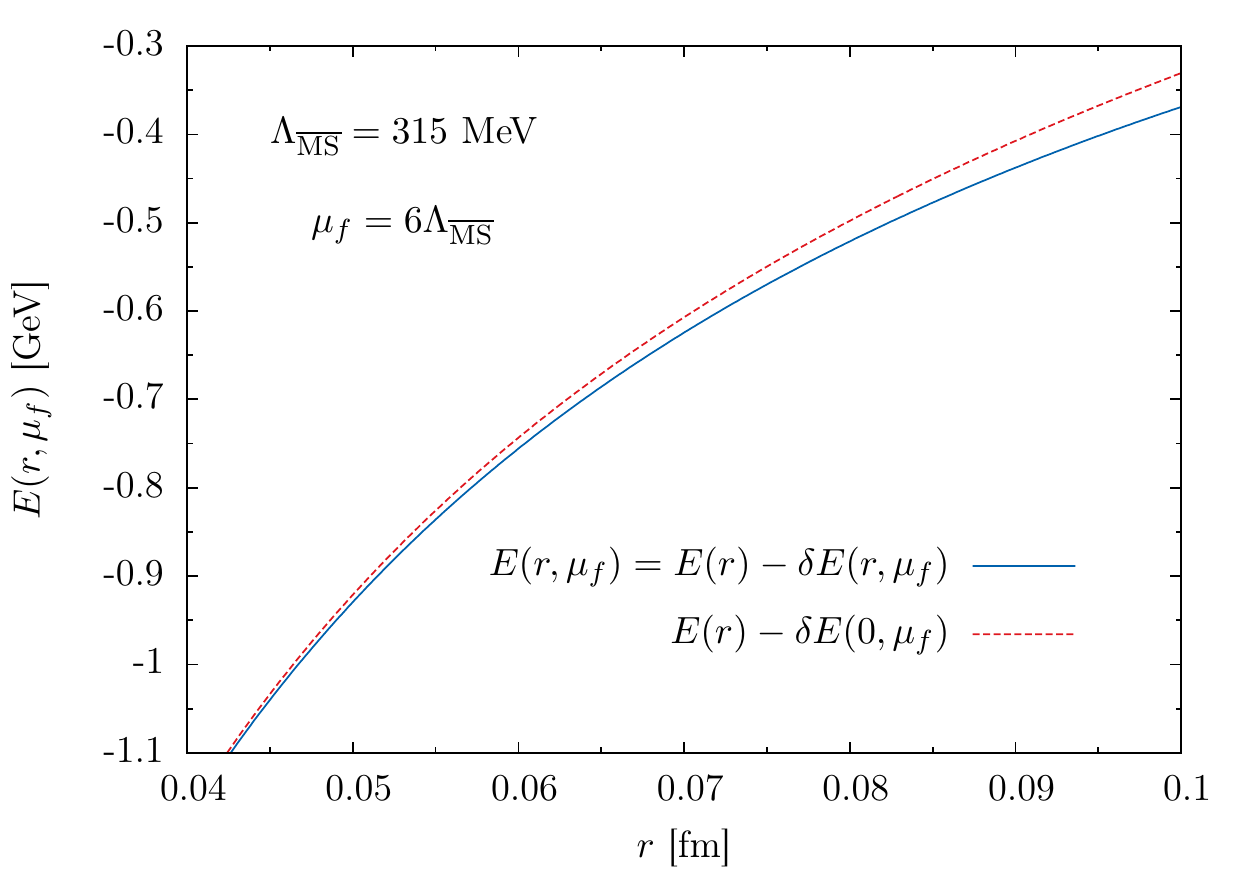} 
\caption{Close-up: Comparison of the static energy as derived by a restricted Fourier transform, obtained by subtracting $\delta E(r,\mu_f)$ in \Eqref{eq:deltaE_expl} from $E(r)$, and
the approximate version, where only the leading, $r$ independent part of $\delta E(r,\mu_f)$ as given by $\delta E(r=0,\mu_f)$ is subtracted; cf. \cite{Beneke:1998rk}. The curves start deviating for $r\gtrsim 0.05\,{\rm fm}$.
This is in accordance with our expectations, as $\delta E(r=0,\mu_f)$ should constitute a good approximation to $\delta E(r,\mu_f)$ for $r^2\mu_f^2\ll1$ only (cf. Appendix~\ref{app:seriesrep});
for $\mu_f=6\Lambda_{\overline{\rm MS}}$ and $r=0.05\,{\rm fm}$ we have $r^2\mu_f^2\approx 0.22$.}
\label{fig:E_bareBenRes}
\end{figure}

For completeness, in Fig.~\ref{fig:E_bareBenRes} we also show a comparison of the static energy~\eqref{eq:pertEres} as obtained from a restricted Fourier transform
and an approximate version, where only the leading, $r$ independent part of $\delta E(r,\mu_f)=\delta E(r=0,\mu_f)+{\cal O}(r^2\mu_f^2)$ is subtracted; see \cite{Beneke:1998rk} and Appendix~\ref{app:seriesrep}.

Let us finally demonstrate the tremendous convergence improvement of the static energy in position space as defined by a restricted Fourier transform in comparison to its definition by a standard unconstrained Fourier transform.
The strongest convergence problems of the perturbative static energy are to be expected for the largest values of $r$ that are accessible within a perturbative approach.
As argued above, the cut-off momentum $\mu_f$ constitutes a viable means to disentangle the perturbative and nonperturbative momentum regimes.
In particular given that the RG momentum scale $\mu$ is identified with $1/r$, this suggested to consider the $r$ interval fulfilling $r\leq1/\mu_f$ as manifestly perturbative only.
This motivates us to quantitatively test the convergence properties of the static energy at the upper bound of the $r$ interval, where the coupling $\alpha_s(1/r)\lesssim1$ reaches its maximum,
and correspondingly set $r=1/\mu_f$. 

\begin{figure}[h]
\center
\includegraphics[width=0.5\textwidth]{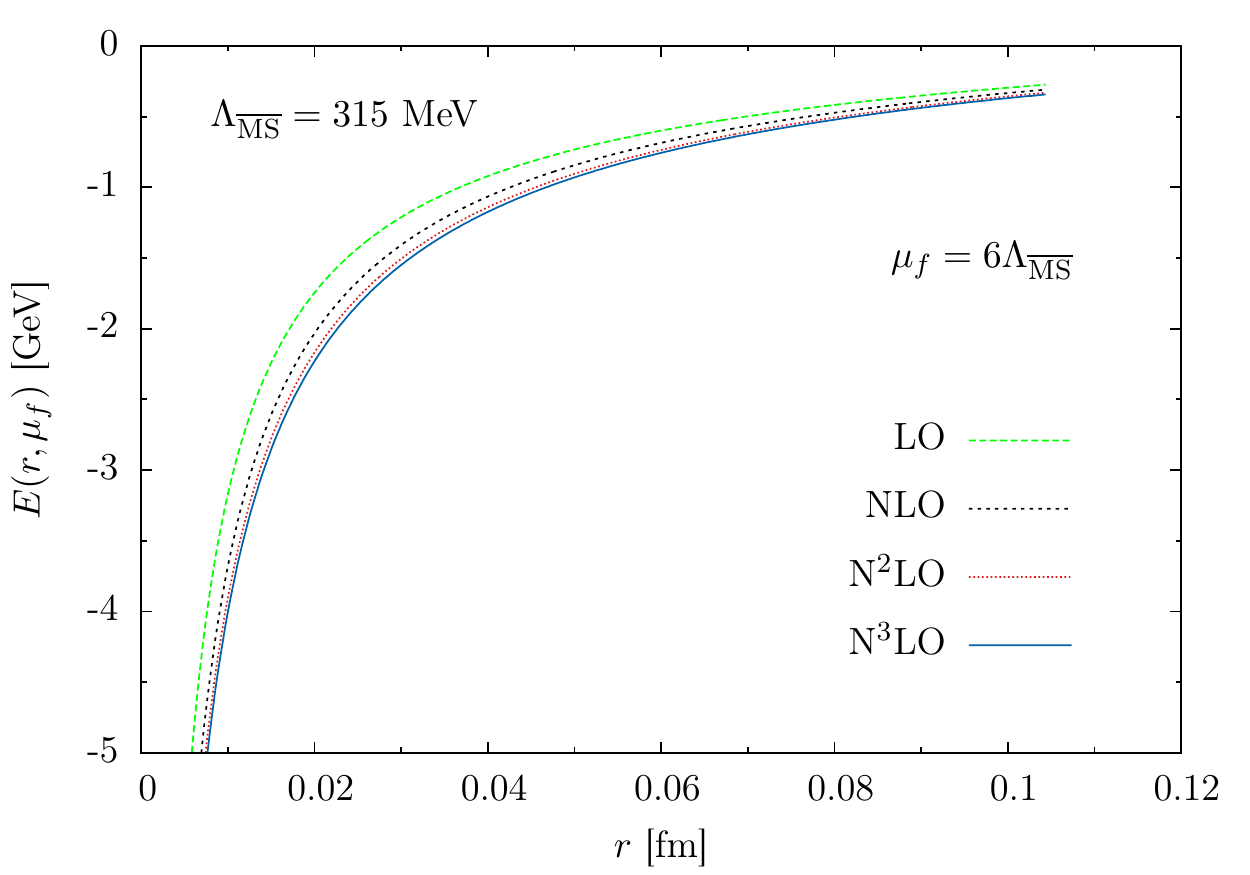} 
\caption{The singlet static energy in position space for $r\leq1/\mu_f$ as obtained by a restricted Fourier transform; $n_f=2$, $\mu=1/r$.
Note the significantly improved convergence properties in contrast to the static energy defined by an unrestricted Fourier transform, as depicted in Fig.~\ref{fig:E0_1234}.
}
\label{fig:E_1234}
\end{figure}

For this purpose it is moreover helpful to introduce some notations: 
The contribution $\sim\alpha_s$ in \Eqref{eq:pertEres} constitutes the leading order (LO) result for the static energy. Terms up to ${\cal O}(\alpha^2_s)$ correspond to next-to-leading order (NLO), and up to ${\cal O}(\alpha^3_s)$ to next-to-next-leading order (N$^2$LO) accuracy. Finally, terms up to ${\cal O}(\alpha^4_s)$ are referred to as N$^3$LO.
Moreover, note that $-E|_{\rm LO}<-E|_{\rm NLO}<-E|_{\rm N^2LO}<-E|_{\rm N^3LO}$ (cf. Figs.~\ref{fig:E0_1234} and \ref{fig:E_1234}).

The results of our examination are shown in Table~\ref{tab:E}.
They clearly confirm that the static energy $E(r,\mu_f)$ as defined by a restricted Fourier transform exhibits very good convergence properties for $r=1/\mu_f$, in the sense that higher order contributions become increasingly less important; see also Fig.~\ref{fig:E_1234}. At the same time, the standard expression of the static energy $E(r)$ as obtained from an unrestricted Fourier transform does not seem to converge at all for this value of $r$; recall Fig.~\ref{fig:E0_1234}.

\begin{table}
 \center
\begin{spacing}{1.5}
\begin{tabular}{c|c|c|}
  & \ unrestricted F.T.\ & \ restricted F.T.\ \\
\hline
 $\frac{{\rm LO}}{{\rm N^3LO}}$ & 38.3\,\% & 92.6\,\% \\
 $\frac{{\rm NLO}-{\rm LO}}{\rm N^3LO}$ & 16.0\,\% & 3.5\,\% \\
 $\frac{{\rm N^2LO}-{\rm NLO}}{\rm N^3LO}$ & 18.7\,\% & 2.2\,\% \\
 $\frac{{\rm N^3LO}-{\rm N^2LO}}{\rm N^3LO}$ & 27.0\,\% & 1.7\,\% \\
\hline
\end{tabular}
\end{spacing}
\caption{This table shows the relative importance of the terms added when improving the accuracy of the static energy $E$ from a given one to the next-better one, normalized to the result for $E$ at the presently best known ($\rm N^3LO$) accuracy.
The second column contains the results for $E(1/\mu_f)$, and the third column those for $E(1/\mu_f,\mu_f)$. The value of $\mu_f$ has been fixed to $\mu_f=6\Lambda_{\overline{\rm MS}}$ ($\Lambda_{\overline{\rm MS}}=315\,{\rm MeV}$), and F.T. stands for Fourier transform.
While for the restricted Fourier transform the newly added terms become increasingly less important towards higher orders, the results for the unrestricted Fourier transform behave uncontrolled. They even show the opposite tendency, i.e., an increase towards higher orders.}
\label{tab:E}
\end{table}

\section{Conclusions} \label{seq:Concl}

In this paper, we have studied in detail the perturbative quark-antiquark static energy in position space as defined by a restricted Fourier transform from momentum space.

Having provided a motivation for the definition of the static energy by means of a restricted Fourier transform in the introduction,
we first summarized the present knowledge of the structure of the static energy in the perturbative regime in Sec.~\ref{seq:basics}.

Here we put special emphasis on its natural decomposition into a strictly perturbative part, made up of contributions $\sim\alpha_s^n$, with $n\in\mathbb{N}$, and a ultrasoft part, which includes terms $\sim\alpha_s^{n+m}\ln^m\alpha_s$, with $n\geq3$ and $m\in\mathbb{N}$ \cite{Brambilla:2006wp,Brambilla:2009bi}.
While the strictly perturbative {\it potential} part is conventionally evaluated order by order in a series expansion in powers of $\alpha_s$ in momentum space,
where it is presently known up to ${\cal O}(\alpha_s^4)$, the corresponding US part is naturally calculated in position space, and starts contributing at ${\cal O}(\alpha_s^4)$.
Of course, this makes the consistent definition of the singlet static energy by means of a restricted Fourier transform significantly more involved as compared to the situation where all contributions are available exclusively in position space. 
It can nevertheless be achieved by a careful analysis of the structure of the US contribution and an adequate choice of the matching coefficients of pNRQCD,
which get modified as compared to the case where the position space potentials emerge by means of a standard (unconstrained) Fourier transforms from their momentum space representations.
Remarkably all relevant integrations necessary to implement the restricted Fourier transform can be performed fully analytically by resorting to just two generic Fourier integrals.

Thereafter, in Sec.~\ref{seq:statEresFT} we have demonstrated in detail that the familiar RG invariance of the static energy in position space up to ${\cal O}(\alpha_s^4)$ as derived by a standard unconstrained Fourier transform is fully retained by its restricted Fourier transform analogue,
the only difference of course being an explicit dependence on the cut-off momentum scale $\mu_f$.
This inherent $\mu_f$ dependence signalizes the limitation of the perturbative expression of the static energy in position space to the manifestly perturbative regime.
Conversely, for the standard unconstrained Fourier transform definition of the perturbative static energy this limitation is only accounted for in the complementary
-- rather vague -- statement that the corresponding result is {\it of course only trustworthy in the perturbative regime}.

Specializing the coupling $\alpha_s$ to its (approximate) $4$-loop solution~\eqref{eq:alpha_s} and promoting the RG scale $\mu$ to $\mu=1/r$, in Sec.~\ref{seq:Ex+Res} we have finally presented some explicit results for the exemplary choice of $n_f=2$ dynamical light flavors. Here our main interest was on the sensitivity of the static energy as defined by the restricted Fourier transform to variations of the momentum cut-off scale $\mu_f$.
Moreover, we highlight the tremendous convergence improvement of the static energy when increasing its accuracy from leading order to the presently best-known ($\rm N^3LO$) accuracy; cf. Fig.~\ref{fig:E_1234}.

Let us finally emphasize again that our analysis facilitates a full analytical determination of the singlet static energy in position space as defined in terms of a restricted Fourier transform at the accuracy with which the perturbative potential in momentum space is known.
Most notably and importantly, this calculation does not involve any additional approximations as compared to the definition by a standard, unrestricted Fourier transform.

It is particularly relevant for comparisons of perturbative calculations and lattice simulations, aiming at the extraction of $\Lambda_{\overline{\rm MS}}$ \cite{Michael:1992nj,Necco:2001gh,Brambilla:2010pp,Jansen:2011vv,Leder:2011pz,Bazavov:2012ka}, and attempts to construct an updated quark-antiquark potential by matching perturbative and lattice results (see, e.g., \cite{Laschka:2011zr}) at the highest possible precision. 

Finally, note that the resummation of US logarithms as performed for the standard definition of the singlet static potential, defined by an unconstrained Fourier transform, in the framework of pNRQCD by \cite{Pineda:2000gza,Brambilla:2009bi}
could also be generalized to its definition by means of a restricted Fourier transform.

\section*{Acknowledgments}

I would like to thank Marc Wagner for numerous stimulating discussions.
Inspiring comments by M. \& M.~Tingu are gratefully acknowledged. Moreover, I am grateful to Karl-Heinz Dubansky for encouragement.

\appendix

\section{Explicit evaluation of the restricted Fourier integral~\eqref{eq:restrictedFT}} \label{app:restrictedFT}

First we substitute the exponential function by its series representation, turn to spherical coordinates and carry out the angle integrations, 
\begin{align}
&\int\!\frac{{d^3 p}}{(2\pi)^3}\,{\rm e}^{i\vec{p}\cdot\vec{r}} \,\frac{4\pi}{p^2}\,L^m\,\Theta(\mu_f-|\vec{p}|) \nonumber \\
 &\quad\quad=\frac{1}{\pi}\int_0^{\mu_f}\!\!dp\,\sum_{n=0}^{\infty}\frac{\left(i pr\right)^n}{n!}\,L^m\int_{-1}^{1}\!d\!\cos\theta \cos^n\theta \nonumber \\
 &\quad\quad=\frac{2}{\pi}\sum_{n=0}^{\infty}\!\frac{(-r^2)^n}{(2n+1)!}\,\!\int_0^{\mu_f}\!\! dp\,p^{2n}\,L^m. \label{eq:step1}
\end{align}

In an intermediate step, we carry out some manipulations of the $p$ integral.  
These manipulations require a specification of the sign of $\ln\left(\mu/\mu_f\right)$.
As emphasized in the second to last paragraph of Sec.~\ref{seq:statEresFT}, in our context the ratio $\mu/\mu_f$ is generically equal or larger than one, such that $\ln\left(\mu/\mu_f\right)\geq0$.

Substituting $p/\mu=e^{-x}$ and subsequently employing formulae 3.381.3 and 3.381.4 of \cite{Gradshteyn} to perform the integral over $x$, we obtain
\begin{multline}
 \int_0^{\mu_f}\!\!{\rm d}p\, p^{2n} L^m=\mu^{2n+1}\int^{\infty}_{\ln(\frac{\mu}{\mu_f})}\!\!{\rm d}x\,(2x)^m{\rm e}^{-(2n+1)x} \\
=\frac{\mu}{2}\,\mu^{2n}\Bigl(\tfrac{2}{2n+1}\Bigr)^{m+1}\Gamma\Bigl(m+1,(2n+1)\ln\frac{\mu}{\mu_f}\Bigr)\,, \label{eq:so1}
\end{multline}
where $\Gamma(\alpha,\chi)$ denotes the incomplete gamma function.
For its first argument being a positive integer, the incomplete gamma function has the following finite sum representation (formula 8.352.2 of \cite{Gradshteyn}),
\begin{equation}
 \Gamma(m+1,\chi)=m!\,{\rm e}^{-\chi}\sum_{j=0}^m\frac{\chi^j}{j!}\,,\quad\text{for}\quad m\in\mathbb{N}_0.
\end{equation}
Inserting this identity in \Eqref{eq:so1},
\begin{align}
 \int_0^{\mu_f}\!\!{\rm d}p\, p^{2n}L^m 
=\mu_f^{2n+1}\sum_{j=0}^m\frac{\frac{m!}{j!}2^{m-j}\ln^j\frac{\mu^2}{\mu_f^2}}{(2n+1)^{m-j+1}}\,, \label{eq:t1}
\intertext{and using a Schwinger parameter integral representation of the factor in the denominator, we arrive at}
=\mu_f\sum_{j=0}^m\!\frac{m!}{j!}\ln^j\frac{\mu^2}{\mu_f^2}\!\int_0^\infty\!\!{\rm d}s\,(r\mu_f{\rm e}^{-s})^{2n}\frac{(2s)^{m-j}{\rm e}^{-s}}{(m-j)!}\,. \label{eq:t2}
\end{align}
Equation~\eqref{eq:t2} constitutes an expansion of \Eqref{eq:so1} in powers of $\ln(\mu^2/\mu_f^2)$.

Plugging \Eqref{eq:t2} into \eqref{eq:step1}, the infinite sum over $n$ can be evaluated explicitly, resulting in
\begin{multline}
\int\!\frac{{d^3 p}}{(2\pi)^3}\,{\rm e}^{i\vec{p}\cdot\vec{r}} \,\frac{4\pi}{p^2}\,L^m\,\Theta(\mu_f-|\vec{p}|)=\frac{2^{m-j+1}}{\pi} \\
\times\sum_{j=0}^{m}\binom{m}{j}\frac{\ln^j\frac{\mu^2}{\mu_f^2}}{r}\!\int_0^\infty\!\!{\rm d}s\,s^{m-j}\sin(r\mu_f{\rm e}^{-s})\,. \label{eq:step3}
\end{multline}

To perform the $s$ integral, we substitute $s$ for $\tau=r\mu_f{\rm e}^{-s}$ and make use of 
$\ln^{m-j}\!\frac{\tau}{r\mu_f}=[\partial_\eta^{m-j}(\frac{\tau}{r\mu_f})^\eta]|_{\eta=0}$ \cite{Peter:1997me}.
This allows us to write
\begin{align}
 &\int_0^\infty\!\!{\rm d}s\,s^{m-j}\sin(r\mu_f{\rm e}^{-s}) \nonumber\\
 &\ =\left.\left[(-\partial_\eta)^{m-j}\int^{r\mu_f}_0\!\!\frac{{\rm d}\tau}{\tau}\,\left(\frac{\tau}{r\mu_f}\right)^\eta\sin(\tau)\right]\right|_{\eta=0} \nonumber\\
 &\ =-\tfrac{r\mu_f}{2}\!\!\left.\left[(-\partial_\eta)^{m-j}\left\{\tfrac{\Gamma(\eta)-\Gamma(\eta,{\rm i}r\mu_f)}{(ir\mu_f)^{1+\eta}}+{\rm c.c.}\right\}\right]\right|_{\eta=0} , \label{eq:int_s}
\end{align}
where we employed formula 3.761.1 of \cite{Gradshteyn} and identity \cite{wolfram:id1} in the last step. The abbreviation ${\rm c.c.}$ stands for complex conjugate.

Combining Eqs.~\eqref{eq:step3} and \eqref{eq:int_s}, we finally obtain \Eqref{eq:restrictedFT}.
Notably, all terms in \Eqref{eq:restrictedFT} can be obtained by parameter differentiations of the master integral expression,
\begin{equation}
 \frac{\Gamma(\eta)-\Gamma(\eta,{\rm i}r\mu_f)}{(ir\mu_f)^{1+\eta}}+{\rm c.c.}\,, \label{eq:masterintexp}
\end{equation}
and are of the generic type
\begin{equation}
 \biggl[\partial_\eta^n\biggl\{\frac{\Gamma(\eta)- \Gamma(\eta,ir\mu_f)}{(ir\mu_f)^{1+\eta}}+{\rm c.c.}\biggr\}\biggr]\bigg|_{\eta=0},\label{eq:diffm_masterintexp}
\end{equation}
with $n\in\mathbb{N}_0$.

\section{Series representation of \Eqref{eq:diffm_masterintexp}} \label{app:seriesrep}

The incomplete gamma function $\Gamma(\alpha,\chi)$ has an exact series representation (formula 8.354.2 of~\cite{Gradshteyn}),
\begin{equation}
 \Gamma(\alpha,\chi)=\Gamma(\alpha)-\sum_{k=0}^\infty\frac{(-1)^k\chi^{\alpha+k}}{k!\,(\alpha+k)}\ ,\ \ \alpha\neq-\mathbb{N}_0\,.
\label{eq:grad1}
\end{equation}
With its help, we write
\begin{align}
&\frac{\Gamma(\eta)- \Gamma(\eta,ir\mu_f)}{(ir\mu_f)^{1+\eta}}+{\rm c.c.} \nonumber\\
&\quad\quad=-2\sum_{k=0}^\infty\frac{(-r^2\mu_f^2)^{k}}{(2k+1)!\,(2k+1+\eta)}\,, \label{eq:grad2a}\\
\intertext{which is also applicable for $\eta=0$, and expressing the last factor in the denominator in terms of a geometric series,}
&\quad=-2\sum_{l=0}^\infty\sum_{k=0}^\infty\frac{(-1)^{l+k}}{(2k+1)!(2k+1)^{1+l}}(r^2\mu_f^2)^{k}\eta^l\,. \label{eq:grad2b}
\intertext{Equation~\eqref{eq:grad2b} constitutes a double expansion in terms of both $r^2\mu_f^2$ and $\eta$.
It can also be expressed as}
&=-2\sum_{l=0}^\infty{}_{l+1}F_{l+2}\Bigl(\!\!\!\underbrace{\tfrac{1}{2},\ldots}_{l+1\ {\rm times},};\underbrace{\tfrac{3}{2},\ldots}_{l+2\ {\rm times}}\!\!;-\tfrac{r^2\mu_f^2}{4}\Bigr)(-\eta)^l,
\end{align}
where ${}_pF_q(\alpha_1,\ldots,\alpha_p;\alpha_2,\ldots,\alpha_q;\chi)$ is the {\it generalized hypergeometric series}; cf. formula 9.14.1 of \cite{Gradshteyn}.
In particular,
\begin{equation}
 {}_1F_2\Bigl(\tfrac{1}{2};\tfrac{3}{2},\tfrac{3}{2};-\tfrac{r^2\mu_f^2}{4}\Bigr)=\frac{{\rm Si}(r\mu_f)}{r\mu_f}\,.
\end{equation}
Thus, for $n\in\mathbb{N}_0$ we obtain
\begin{multline}
\biggl[\partial_\eta^n\biggl\{\frac{\Gamma(\eta)- \Gamma(\eta,ir\mu_f)}{(ir\mu_f)^{1+\eta}}+{\rm c.c.}\biggr\}\biggr]\bigg|_{\eta=0} \\
=\sum_{l=0}^\infty d_l(n)(r^2\mu_f^2)^{l}, \label{eq:diffm_masterintexp1}
\end{multline}
where we defined
\begin{equation}
d_l(n)\equiv\frac{2n!(-1)^{1+n+l}}{(2l+1)!(2l+1)^{1+n}} \label{eq:dk}
\end{equation}
for future reference.
Equation~\eqref{eq:diffm_masterintexp1} constitutes the desired series representation of \Eqref{eq:diffm_masterintexp}.
Alternatively, it can be represented as
\begin{multline}
\biggl[\partial_\eta^n\biggl\{\frac{\Gamma(\eta)- \Gamma(\eta,ir\mu_f)}{(ir\mu_f)^{1+\eta}}+{\rm c.c.}\biggr\}\biggr]\bigg|_{\eta=0} \\
=2n!(-1)^{n+1}{}_{n+1}F_{n+2}\Bigl(\!\!\!\underbrace{\tfrac{1}{2},\ldots}_{n+1\ {\rm times},};\underbrace{\tfrac{3}{2},\ldots}_{n+2\ {\rm times}}\!\!;-\tfrac{r^2\mu_f^2}{4}\Bigr). \nonumber
\end{multline}
In the limit $r\mu_f=0$ only the $l=0$ term in \Eqref{eq:diffm_masterintexp1} contributes, such that
\begin{equation}
 \eqref{eq:diffm_masterintexp1} \quad\to\quad d_0(n)=2n!(-1)^{1+n}\,. \label{eq:d0}
\end{equation}

\section{A closer look on the series representation~\eqref{eq:diffm_masterintexp1}} \label{app:closerlook}

Now we briefly investigate the scaling behavior of the expansion coefficients $d_n(l)$ in \Eqref{eq:diffm_masterintexp1}.

The scaling of $d_0(n)$ with respect to $n$ can be read off \Eqref{eq:d0}. Its modulus grows factorial with $n$, and it is straightforward to see that
\begin{equation}
 d_0(n+1)=-(n+1)\,d_0(n)\,. \label{eq:d0_ind}
\end{equation}

In order to simplify the discussion for larger values of $l$ and $n$, we make use of Stirling's formula,
\begin{equation}
 n!=\sqrt{2\pi n}\left(\frac{n}{\rm e}\right)^n\left(1+\frac{1}{12n}+{\cal O}\bigl(\tfrac{1}{n^2}\bigr)\right), \label{eq:Stirling}
\end{equation}
with ${\rm e}=\exp(1)$, to express the factorials contained in \Eqref{eq:dk} in terms of powers.
For the rather rough estimates we are interested in in the following it is basically fine to adopt this approximation for all values of $\{n,l\}\geq1$ [Note the prefactor $\frac{1}{12}$ of the $\frac{1}{n}$ correction in \Eqref{eq:Stirling}.].

This results in
\begin{multline}
 d_l(n)=\frac{2}{\sqrt{\rm e}}(-1)^{1+n+l}\left(\frac{{\rm e}}{2l+1}\right)^{2(l+1)} \\
 \times\left(\frac{n}{(2l+1){\rm e}}\right)^{n+\frac{1}{2}}\left(1+{\cal O}\bigl(\tfrac{1}{n}\bigr)+{\cal O}\bigl(\tfrac{1}{2l+1}\bigr)\right), \label{eq:dk_asympt}
\end{multline}
and,
employing
\begin{equation}
 \left(\frac{n+1}{n}\right)^{n+\frac{1}{2}}={\rm e}+{\cal O}\bigl(\tfrac{1}{n^2}\bigr), \quad {\cal O}\bigl(\tfrac{1}{n+1}\bigr)={\cal O}\bigl(\tfrac{1}{n}\bigr),
\end{equation}
implies
\begin{equation}
 d_l(n+1)=-\frac{n+1}{2k+1}\,d_l(n). \label{eq:dk_ind}
\end{equation}
Thus, in comparison to \Eqref{eq:d0_ind} the ratio of the coefficients $d_l(n+1)$ and $d_l(n)$ is diminished by an overall factor of $\sim\frac{1}{2l+1}$.
From \Eqref{eq:dk_ind} we infer that $|d_l(n+1)|>|d_l(n)|$ for $n>2l$.

Finally, one may ask how large $n$ has to be for given $l\geq1$ such that $|d_l(n)|>1$.
Equation~\eqref{eq:dk_asympt} implies that this should be the case for {\it for sufficiently large} $n$. More precisely, 
specializing $n$ to $n_0(l)\equiv(2l+1)^2$ the power-like decrease with $\frac{e}{2l+1}$ can be compensated by powers of $\frac{n}{(2l+1)e}$, yielding
\begin{equation}
 |d_l(n_0)|=\frac{2}{\sqrt{\rm e}}\left(\!\frac{\sqrt{n_0}}{{\rm e}}\right)^{n_0-\sqrt{n_0}}
 \!\left(1+{\cal O}\bigl(\tfrac{1}{\sqrt{n_0}}\bigr)\right)>1, \label{eq:cond2}
\end{equation}
such that $d_l(n)>1$ for at least $n\geq n_0(l)$.

\end{document}